\documentclass[aps,prd,eqsecnum,showpacs,amsmath,nofootinbib,superscriptaddress,twocolumn,floats,preprintnumbers]{revtex4}

\usepackage[dvips]{epsfig,color,graphicx}
\usepackage{amsfonts,amssymb,theorem,mathrsfs,times}
\definecolor{red  }{rgb}{1,0,0}
\definecolor{blue }{rgb}{0,0,1}
\definecolor{green}{rgb}{0,1,0}
\newcommand{\mn}{{\mu\nu}}

\renewcommand{\a}{\alpha}
\newcommand{\ta}{\tilde{\a}}

\renewcommand{\c}{\gamma}
\renewcommand{\d}{\delta}
\newcommand{\e}{\epsilon}
\renewcommand{\k}{\kappa}
\newcommand{\s}{\sigma}

\newcommand{\la}{\lambda}

\newcommand{\pa}{\partial}
\newcommand{\na}{\nabla}

\newcommand{\nn}{\nonumber \\}
\newcommand{\lh}{\left(}
\newcommand{\rh}{\right)}

\newcommand{\hg}{\hat{g}}
\newcommand{\hl}{\hat{\la}}
\newcommand{\hr}{\hat{r}}
\newcommand{\hf}{\hat{f}}
\newcommand{\hh}{\hat{h}}

\newcommand{\hp}{\hat{\phi}}
\newcommand{\hm}{\hat{\mu}}
\newcommand{\hn}{\hat{\nu}}
\newcommand{\hx}{\hat{\xi}}

\newcommand{\p}[1]{(\ref{#1})}

\newcommand{\dalm}{\kern1pt\vbox{\hrule height 0.9pt\hbox{\vrule width
0.9pt\hskip 2.5pt\vbox{\vskip 5.5pt}\hskip 3pt\vrule width 0.3pt}\hrule height
0.3pt}\kern1pt}

\usepackage{amsmath}	% required for `\align' (yatex added)
\begin{document}

\preprint{WU-AP/307/09}
\preprint{KU-TP 034}

%<<<<<<<<<<<<< TITLE >>>>>>>>>>>>>>>%
%%%%%%%%%%%%%%%%%%%%%%%%%%%%%%%%%%%%%%%%%%%%%%%%%%%%%%%%%%%%%%%%%%
%%%%%%%%%%%%%%%%%%%%%%%%%%%%%%%%%%%%%%%%%%%%%%%%%%%%%%%%%%%%%%%%%%
\title{Black Hole Solutions in String Theory with
Gauss-Bonnet Curvature Correction
}
%%%%%%%%%%%%%%%%%%%%%%%%%%%%%%%%%%%%%%%%%%%%%%%%%%%%%%%%%%%%%%%%%%
%%%%%%%%%%%%%%%%%%%%%%%%%%%%%%%%%%%%%%%%%%%%%%%%%%%%%%%%%%%%%%%%%%

%<<<<<<<<<<<<< AUTHOR >>>>>>>>>>>>>>>%
%<<<<<<<<<<<<< MAIL >>>>>>>>>>>>>>>%
%<<<<<<<<<<<<< AFFILIATION >>>>>>>>>>>>>>>%
\author{Kei-ichi Maeda}
\email{maeda@waseda.jp}
\affiliation{
Department of Physics, Waseda University,
Shinjuku, Tokyo 169-8555, Japan
}
\affiliation{
Advanced Research Institute for Science and Engineering,
Waseda University, Shinjuku, Tokyo 169-8555, Japan
}
\author{Nobuyoshi Ohta}
\email{ohtan@phys.kindai.ac.jp}
\affiliation{
Department of Physics, Kinki University, Higashi-Osaka, Osaka 577-8502,
Japan
}
\author{Yukinori Sasagawa}
\email{yukinori@gravity.phys.waseda.ac.jp}
\affiliation{
Department of Physics, Waseda University,
Shinjuku, Tokyo 169-8555, Japan
}

%<<<<<<<<<<<<< DATE >>>>>>>>>>>>>>>%
\date{\today}

%======================================%
%<<<<<<<<<<<<< ABSTRACT >>>>>>>>>>>>>>>%
%======================================%
%%%%%%%%%%%%%%%%%%%%%%%%%%%%%%%%%%%%%%%%%%%%%%%%%%%%%%%%%%%%%%%%%%
%%%%%%%%%%%%%%%%%%%%%%%%%%%%%%%%%%%%%%%%%%%%%%%%%%%%%%%%%%%%%%%%%%
\begin{abstract}
We present the black hole solutions and analyse their properties
 in the superstring effective field
theory with the Gauss-Bonnet curvature correction terms.
We find qualitative differences in our results from those obtained in
the truncated model in the Einstein frame.
The main difference in our model from the truncated one
is that the existence of a turning point in the mass-area curve,
the mass-entropy curve, and the mass-temperature curve in
five and higher dimensions, where we expect a change of stability.
We also find a mass gap in our model, where there is no black hole solution.
In five dimensions, there exists
a maximum black hole temperature and the temperature vanishes
at the minimum mass, which is not found in the truncated model.
\end{abstract}
%%%%%%%%%%%%%%%%%%%%%%%%%%%%%%%%%%%%%%%%%%%%%%%%%%%%%%%%%%%%%%%%%%
%%%%%%%%%%%%%%%%%%%%%%%%%%%%%%%%%%%%%%%%%%%%%%%%%%%%%%%%%%%%%%%%%%

%<<<<<<<<<<<<< PACS NUMBER >>>>>>>>>>>>>>>%
\pacs{
04.20.Cv, %Fundamental problems and general formalism
% 04.20.Ha,%Asymptotic structure
04.50.-h,%Higher-dimensional gravity and other theories of gravity
% 04.50.+h, %Gravity in more than four dimensions, Kaluza-Klein theory,
	  %unified field theories; alternative theories of gravity
04.60.Cf, %Gravitational aspects of String theory
04.70.Dy. %Quantum aspects of black holes, evaporation, thermodynamics
}

\maketitle

%======================================%
%<<<<<<<<<<<< SECTION I  >>>>>>>>>>>>>>%
%======================================%
%%%%%%%%%%%%%%%%%%%%%%%%%%%%%%%%%%%%%%%%%%%%%%%%%%%%%%%%%%%%%%%%%%%
%%%%%%%%%%%%%%%%%%%%%%%%%%%%%%%%%%%%%%%%%%%%%%%%%%%%%%%%%%%%%%%%%%%
%%%%%%%%%%%%%%%%%%%%%%%%%%%%%%%%%%%%%%%%%%%%%%%%%%%%%%%%%%%%%%%%%%%
\section{Introduction}
%%%%%%%%%%%%%%%%%%%%%%%%%%%%%%%%%%%%%%%%%%%%%%%%%%%%%%%%%%%%%%%%%%%
%%%%%%%%%%%%%%%%%%%%%%%%%%%%%%%%%%%%%%%%%%%%%%%%%%%%%%%%%%%%%%%%%%%
%%%%%%%%%%%%%%%%%%%%%%%%%%%%%%%%%%%%%%%%%%%%%%%%%%%%%%%%%%%%%%%%%%%
A black hole always absorbs the ambient matter and
the mass increases in time classically.
However, if we take into account the quantum effect,
a black hole will emit the Hawking radiation and evaporate away.
It behaves as a thermal object.
Of course, such a quantum effect can be ignored
for astrophysical black holes.
On the other hand, for microscopic black holes,
the situation drastically changes.
The black hole loses its mass by the radiation and may vanish.
When the mass approaches the Planck mass, however,
the semiclassical approach is no longer valid.
To know what happens at the end of evaporation, i.e.,
to answer the questions such as
 ``What is the final state of black hole?"
or ``Does a naked singularity appear?",
we may need to study it by quantum gravity.

One of the most promising candidates for the quantum theory of
gravity is the string theory, which may also provide us a unified theory
of fundamental interactions (the so-called ``theory of everything")~\cite{string}.
String theory is, however, still in a developing stage
and may not yet be able to treat strong gravitational phenomena
such as a black hole directly. Hence we shall study black holes in the effective
field theory including string quantum correction terms.
The field theory limit of the string theories
leads to a ten-dimensional supergravity theory at the lowest derivative level.
In addition, it is known that quantum effect gives higher
curvature correction terms.
There are five string theories in ten dimensions,
which are related with each other via dualities.
The curvature correction terms depend on the type of string theory.
In the heterotic string theory, the lowest corrections are described by
the second-order curvature term, i.e., the so-called
Gauss-Bonnet term~\cite{R2_correction}.
On the other hand, in type II string theory, the fourth-order
curvature terms appear as the lowest~\cite{R4_correction}.

The Gauss-Bonnet term is known as the second-order Lovelock gravity.
The Lovelock theory is a higher curvature generalization of Einstein gravity.
Its field equations contain terms up to the second-order derivatives
of the metric functions and the second-order derivative terms are
linear~\cite{Lovelock,Zumino}.
The $n$-th order of Lovelock gravity is constructed by the Euler density in
the $2n$ dimensional spacetime.
Hence, $n$-th terms with $n\leq [(D-1)/2]$ contribute to the field equations.
The black hole solution in the theory with the Gauss-Bonnet term or with the Lovelock
action has been analysed in the models  in many works~\cite{BH_EGB,BH_Lovelock}.

A dilaton field also plays an important role in the string theory
as a dynamical field. Hence dilatonic models have been studied intensively
in the context of string theory.
The black hole solution in such a dilatonic theory
was studied in~\cite{GM} and~\cite{GHS}.
The black hole solution with the Riemann curvature squared correction term
coupled to dilaton was first studied by the linear perturbation
approach~\cite{Callan}.
When the Gauss-Bonnet term couples to a dilaton,
it contributes to the dynamical equations even in four dimensions.
Full study of this case requires numerical evaluation, and has been made
in~\cite{degb,degb2,Torii_Yajima_Maeda,Chen1,Chen2,ohta_torii1}.

In the string frame, the Einstein-Hilbert curvature term is also coupled to
the dilaton field. So usually we perform a conformal transformation
to find the Einstein frame, in which the Einstein-Hilbert
curvature term does not couple to the dilaton field.
When we study a black hole in the Einstein frame, only the Gauss-Bonnet
term is taken into account as the quantum correction,
but some additional terms appear through a conformal transformation
compared with the string frame.
It is not so obvious which frame is to be used in investigating solutions
where strong gravitational effects become strong such as black holes.
It is certainly natural to take the action in the string frame in string theory,
and then it is important to check if the above additional terms make any
difference in the results.
In this paper, we analyse black hole solutions in the effective action
in the Einstein frame equivalent to that in the string frame,
and compare the results with those in the truncated
effective action, i.e., the model only with the Gauss-Bonnet term
as the correction.
Black hole solutions in the four-dimensional string frame are examined in the context of
black hole - string transition in~\cite{Davydov}.

This paper is organized as follows:
In Sec.~\ref{sec2}, we
present the effective action which we discuss in this paper,
and perform a conformal transformation to obtain the description in
the Einstein frame. We also define our truncated model.
In Sec.~\ref{sec3}, we write down the basic equations for
a spherically symmetric and static spacetime in the dilatonic
Einstein-Gauss-Bonnet theory, and give the boundary condition
for the regular black hole horizon.
We transform the variables in the string frame to
those in the Einstein frame in Sec.~\ref{sec4}.
In Sec.~\ref{sec5}, we introduce the thermodynamical variables.
We then show our numerical results in  Sec.~\ref{sec6}.
In Sec.~\ref{sec7}, we briefly summarize the truncated dilatonic
Einstein-Gauss-Bonnet model.
We present the basic equations and the boundary conditions on the
black hole horizon.
We compare our results in the dilatonic
Einstein-Gauss-Bonnet theory with those in the truncated one in
Sec.~\ref{sec8}.
The concluding remarks are made in Sec.~\ref{sec9}.

%%%%%%%%%%%%%%%%%%%%%%%%%%%%%%%%%%%%%%%%%%%%%%%%%%%%%%%%%%%%%%%%%%%
%%%%%%%%%%%%%%%%%%%%%%%%%%%%%%%%%%%%%%%%%%%%%%%%%%%%%%%%%%%%%%%%%%%
%%%%%%%%%%%%%%%%%%%%%%%%%%%%%%%%%%%%%%%%%%%%%%%%%%%%%%%%%%%%%%%%%%%
\section{Effective action and its truncation in the Einstein frame}
\label{sec2}
%%%%%%%%%%%%%%%%%%%%%%%%%%%%%%%%%%%%%%%%%%%%%%%%%%%%%%%%%%%%%%%%%%%
%%%%%%%%%%%%%%%%%%%%%%%%%%%%%%%%%%%%%%%%%%%%%%%%%%%%%%%%%%%%%%%%%%%
%%%%%%%%%%%%%%%%%%%%%%%%%%%%%%%%%%%%%%%%%%%%%%%%%%%%%%%%%%%%%%%%%%%
In this paper, we focus on the Einstein-Gauss-Bonnet gravity
coupled to a dilaton field.
%%%%%%%%%%%%%%%%%%%%%%%%%%%%%%%%%%%%%%%%%%%%%%%%%%%%%%%%%%%%%%%%%%%
%%%%%%%%%%%%%%%%%%%%%%%%%%%%%%%%%%%%%%%%%%%%%%%%%%%%%%%%%%%%%%%%%%%
%\subsection{String frame v.s. Einstein frame}
%%%%%%%%%%%%%%%%%%%%%%%%%%%%%%%%%%%%%%%%%%%%%%%%%%%%%%%%%%%%%%%%%%%
%%%%%%%%%%%%%%%%%%%%%%%%%%%%%%%%%%%%%%%%%%%%%%%%%%%%%%%%%%%%%%%%%%%
The effective action of the heterotic string theory
in the string frame is given by
\begin{align}
{\cal S}_{\rm S} =\frac{1}{2\k_{D}^2} \int d^D \!\hat x \sqrt{-\hg} \ e^{-2\hp}
 \biggl(\hat R +4(\hat{\na} \hp)^2
 +  \a_{2} \hat R_{GB}^2 \biggr),
\label{Saction}
\end{align}
where $\k_{D}^2$ is the $D$-dimensional gravitational constant,
$\hat \phi$ is a dilaton field,
\begin{align}
\hat R_{GB}^2  = \hat R^2 -4 \hat R_{\mn} \hat R^{\mn}
+\hat R_{\mn\rho\s}\hat R^{\mn\rho\s},
\end{align}
is the Gauss-Bonnet curvature term, and $\a_2=\alpha'/8$ is its coupling
constant.

In the string frame, the dilaton field couples to the Ricci scalar curvature
nonminimally. Hence we perform a conformal transformation
\begin{eqnarray}
g_\mn = \exp[-2\c^2\hp]\, \hg_\mn
,
\label{conf_trans}
\end{eqnarray}
where
$\c^2=\frac{2}{D-2}$, in order to find the Einstein-Hilbert action.
The action in the Einstein frame is given by
\begin{eqnarray}
{\cal S}_{\rm E} &=&\frac{1}{2\k_{D}^2} \int d^D \!x \sqrt{-g} \
 \biggl[ R -\frac{1}{2}
(\na  \phi)^2
\nn
& +&
\a_{2} e^{-  \c\phi}  \left(R_{GB}^2 +
 {\cal F}[\na\phi , R]\right) \biggr]\,,
\label{Eaction}
\end{eqnarray}
where we have introduced $\phi = 2 \c \hp$~\cite{footnote1}.
$R$ and $R_{GB}^2$ are the Ricci scalar curvature and the Gauss-Bonnet
curvature term with respect to the Einstein-frame
metric $g_{\mn}$, respectively.
Because we have the Gauss-Bonnet term, if we start from
the effective action in the string frame,
there appears the additional complicated term ${\cal F}$
in the Einstein frame, which is given by
\begin{widetext}
\begin{eqnarray}
{\cal F}[\na\phi , R]
&=&4(D-3) \c R_{\mu\nu}  \nabla^\mu\nabla^\nu\phi
 - 2(D-3)\c^2 R_{\mu\nu} (\nabla^\mu\phi) (\nabla^\nu\phi)
- 2(D-3)\c R \nabla^2\phi
-\frac12 (D-3)_4 \c^2 R (\nabla \phi)^2
\nn
&&
- (D-2)_3 \c^2(\nabla_\mu\nabla_\nu\phi)^2
+ (D-2)_3 \c^2(\nabla^2\phi)^2
+(D-2)_3 \c^3 (\nabla^{\mu}\phi)(\nabla^{\nu}  \phi)
(\nabla_{\mu}\nabla_{\nu}\phi)
\nn
&&
+\frac12 (D-2)(D-3)^2 \c^3 (\nabla^2\phi)(\nabla\phi)^2
+ \frac{1}{16}(D-1)_4 \c^4 \left[(\nabla\phi)^2\right]^2
\nonumber \\
&=&D(D-3) \c^2 G_{\mu\nu}  \nabla^\mu\phi\nabla^\nu\phi
 +\frac12 (D-1)_3 \c^3 (\nabla^2\phi)(\nabla\phi)^2
+ \frac{1}{16}(D-1)_4 \c^4 \left[(\nabla\phi)^2\right]^2
+({\rm surface~term})
\,,
~~~
\label{extra_terms}
\end{eqnarray}
\end{widetext}
where $G_{\mu\nu}$ is the Einstein tensor.
Here we have used a concise notation
\begin{eqnarray}
(D-m)_n:= (D-m)(D-m-1)\cdots (D-n)\,,
\end{eqnarray}
with $m$ and $n$ being some integers ($n>m$).

This term ${\cal F}$ has not been sometimes considered
in many literatures when cosmology or black hole solutions are studied
in the Einstein frame~\cite{degb,degb2,Torii_Yajima_Maeda,Chen1,Chen2,ohta_torii1}.
However, if we start with the effective action in the string frame~(\ref{Saction}),
there exists the complicated term~\p{extra_terms} because the action with ${\cal F}$
in the Einstein frame (\ref{Eaction}) is classically equivalent to the original one.
We note that this does not mean that the theory in the Einstein frame without ${\cal F}$
is not correct, but there is an intrinsic ambiguity in the theory.
So it is interesting and important to study if this makes any difference
in the obtained results.

In order to see the effects of this extra term ${\cal F}$,
we study a black hole solution in this paper.
For this purpose, we solve two sets of equations; one is a set of
equations including the ${\cal F}$ term, and the other is that without
the ${\cal F}$ term. We shall call the former and the latter
the dilatonic Einstein-Gauss-Bonnet (DEGB)
theory and  the truncated one (TDEGB), respectively.
The action for TDEGB theory is given by
\begin{eqnarray}
{\cal S}_{\rm T}
&=&\frac{1}{2\k_{D}^2} \int d^D \!x
\sqrt{-g}
 \biggl[ R -\frac{1}{2}(\na \phi)^2
+ \a_{2} e^{-  \c\phi}  R_{GB}^2  \biggr]
\,,
\nn
&&
~~
\label{Taction}
\end{eqnarray}
Although one can solve the basic equations of the DEGB theory in the Einstein
frame~(\ref{Eaction}), it is easier to solve them in the string frame~(\ref{Saction})
and to transform the solutions in the string frame
into those in the Einstein frame by the conformal transformation
(\ref{conf_trans}). This is the strategy we take here.

%%%%%%%%%%%%%%%%%%%%%%%%%%%%%%%%%%%%%%%%%%%%%%%%%%%%%%%%%%%%%%%%%%%
%%%%%%%%%%%%%%%%%%%%%%%%%%%%%%%%%%%%%%%%%%%%%%%%%%%%%%%%%%%%%%%%%%%
%%%%%%%%%%%%%%%%%%%%%%%%%%%%%%%%%%%%%%%%%%%%%%%%%%%%%%%%%%%%%%%%%%%
\section{Dilatonic Einstein-Gauss-Bonnet model in the string frame}
\label{sec3}
%%%%%%%%%%%%%%%%%%%%%%%%%%%%%%%%%%%%%%%%%%%%%%%%%%%%%%%%%%%%%%%%%%%
%%%%%%%%%%%%%%%%%%%%%%%%%%%%%%%%%%%%%%%%%%%%%%%%%%%%%%%%%%%%%%%%%%%
%%%%%%%%%%%%%%%%%%%%%%%%%%%%%%%%%%%%%%%%%%%%%%%%%%%%%%%%%%%%%%%%%%%

%%%%%%%%%%%%%%%%%%%%%%%%%%%%%%%%%%%%%%%%%%%%%%%%%%%%%%%%%%%%%%%%%%%
%%%%%%%%%%%%%%%%%%%%%%%%%%%%%%%%%%%%%%%%%%%%%%%%%%%%%%%%%%%%%%%%%%%
\subsection{Basic equations in the string frame}
%%%%%%%%%%%%%%%%%%%%%%%%%%%%%%%%%%%%%%%%%%%%%%%%%%%%%%%%%%%%%%%%%%%
%%%%%%%%%%%%%%%%%%%%%%%%%%%%%%%%%%%%%%%%%%%%%%%%%%%%%%%%%%%%%%%%%%%
First we present the basic equations in the string frame.
To find a black hole solution, we assume a spherically symmetric and
static spacetime, whose metric form is given by
\begin{align}
d\hat{s}^2_D=-e^{2\hn}dt^2+e^{2\hl}d\hr^2+e^{2\hm} d\Omega_{D-2}^2 ,
\label{metric_string}
\end{align}
where $\hn,\hl$, and $\hm$ are functions of the radial coordinate $\hat r$.
$d\Omega_{D-2}^2$ is the metric of $(D-2)$-dimensional unit sphere.
We derive the explicit form of the action with this ansatz as
\begin{widetext}
\begin{align}
{\cal S}_{\!S}=
\frac{1}{2\k_{D}^2} \int d^D \!x e^{\hat{W}-2\hp}
 \biggl\{ & e^{-2\hl}\bigl[ 2 (D-2) \hat{Y} +(D-2)_3
\hat{A} -4 \hp' \hn' \bigr]
 \biggr.
 +4 e^{-2\hl} \hp'^2 \nn
 + &  \ta_2 e^{-4\hl}
 \bigl[ (D-4)_5 \hat{A}^{2} +4 (D-4) \hat{Y} \hat{A}
 -8 \hp' \hn' \hat{A}
\bigr] \biggl. \biggr\},
\end{align}
\end{widetext}
where we have introduced three variables
$\hat{Y}, \hat{A}$, and $\hat{W}$ by
\begin{eqnarray}
&&\hat{Y}= ~  -\left(\hm''+\hm'^2 -\hm' \hl'\right) \,, ~~
\nn
&&\hat{A}= ~  e^{ 2 ( \hl -\hm )} -\hm'^2  \,,
\nn
&&\hat{W}= ~  \hn +\hl +(D-2)\hm \,,
\end{eqnarray}
and
dropped the surface term.
A prime denotes a derivative with respect to $\hr$.
For brevity, we have introduced the rescaled coupling
constant as $\ta_2 := (D-2)_3 \a_2$, and in what follows,
we will normalize the variables by it.
Taking variations of the action
with respect to $\hp, \ \hn, \ \hl$, and $\hm$,
we find the basic equations.
Since we are interested in a black hole solution,
it may be convenient to introduce new metric functions
$f$ and $\delta$ as
\begin{gather}
ds^2 = -\hat f(\hat r) e^{-2\hat \d(\hat r)} dt^2
+{1\over \hat f(\hat r)} d\hat r^2 +\hat r^2 d\Omega_{D-2}^2,
\label{metric_string2}
\end{gather}
Here we have fixed one metric component as
$e^{2\hat \mu}=\hat r^2$ by using the gauge freedom.
Using the new variables and defining the following variables by
\begin{eqnarray}
\hat h&:= & \hat r(\hat f'-2\hat f \hat \delta')
\,,
\\
\hat X& := & \frac{1}{4\hat f^2
\hat r^2} \left[ \hat h(\hat f'\hat r-\hat h) -2\hat f (\hat h'
\hat r - \hat h ) \right]
\,,
\end{eqnarray}
the basic equations are written as
\begin{widetext}
\begin{align}
\hat f \hat r^4
 F_{{\rm S}(\hp)}&:= -2\Bigl\{ \hr^2 \bigl[ (D-2)_{3} (1-\hf) -(D-2)
(\hf'\hr+\hh) +2 \hat X \hf \hr^2
 +2 (\hf'\hr+\hh) \hp' \hr +4(D-2) \hf \hp'\hr + 4\hf ( \hp'' - \hp'^2 )
\hr^2 \bigr] \nn
&\hspace{0.7cm}+ \ta_2
 \bigl[
(D-4)_{5} (1-\hf)^2 -2 (D-4) (1-\hf) (\hf'\hr+\hh) +4
\hat X (1-\hf) \hf \hr^2
+ 2 \hh \hf' \hr
 \bigr] \Bigr\} =0,
\label{eq_phi}\\
\hat f \hat r^4 F_{{\rm S}(\hn)}&:=
\hr^2 \bigl[(D-2)_3 (1-\hf)
 -(D-2) (\hf' - 4\hf \hp' ) \hr +4 \hf ( \hp'' -\hp'^2 ) \hr^2
 +2 \hp' \hf' \hr^2 \bigr] \nn
&+ \ta_2
 \bigl[
(D-4)_{5} (1-\hf)^2 -2 (D-4) (1-\hf) (\hf' - 4\hf \hp' ) \hr
\nn
& \hspace{1.6cm}
+8  \hf (1-\hf) ( \hp'' -2 \hp'^2 ) \hr^2 +4 (1-3\hf) \hp' \hf' \hr^2
 \bigr] =0,
\label{eq_nu}\\
\hat f \hat r^4  F_{{\rm S}(\hl)}&:=
\hr^2 \bigl[(D-2)_3 (1-\hf) -(D-2) ( \hh -4 \hf \hp' \hr )
 +2 ( \hh -2 \hf \hp' \hr ) \hp' \hr \bigr] \nn
&+ \ta_2
 \bigl[
(D-4)_{5} (1-\hf)^2 -2 (D-4) (1-\hf) ( \hh -4 \hf \hp' \hr )
 +4 (1-3\hf) \hh \hp' \hr
 \bigr] =0,
\label{eq_lam}\\
\hat f \hat r^4 F_{{\rm S}(\hm)}&:=
\hr^2 \bigl[ (D-2)_4 (1-\hf) -(D-2)_3 (\hf'\hr+\hh) +4 (D-2)_3 \hf \hp' \hr
 +2 (D-2) \hat X \hf \hr^2 \nn
& \hspace{1.6cm} +4(D-2) \hf ( \hp'' - \hp'^2 ) \hr^2
 + 2 (D-2) (\hf'\hr+\hh) \hp' \hr \bigr] \nn
&+ \ta_2
 \bigl[
(D-4)_{6} (1-\hf)^2 -2 (D-4)_5 (1-\hf) (\hf'\hr+\hh) +8 (D-4)_5 (1-\hf)
\hf \hp' \hr
\nn
& \hspace{1.6cm}
 +4(D-4) \hat X (1-\hf) \hf \hr^2
+8(D-4) \hf (1-\hf) ( \hp'' - 2 \hp'^2 ) \hr^2
 +4(D-4) (1-3\hf) (\hf'\hr+\hh) \hp' \hr \nn
& \hspace{1.6cm} +2 (D-4) \hh \hf' \hr -8 \hf \hh ( \hp'' - 2 \hp'^2 ) \hr^2
 -8 \hh \hf' \hp' \hr^2 +16 \hat X \hp' \hf^2 \hr^3
 \bigr] =0
\,. \label{eq_mu}
\end{align}
\end{widetext}
Because of the Bianchi identity, there is one relation between the above
four functionals; $ F_{{\rm S}(\hp)},  F_{{\rm S}(\hn)},
 F_{{\rm S}(\hl)}$, and $ F_{{\rm S}(\hm)}$, i.e.,
\begin{eqnarray}
&&
\hat f^{-1/2}\left(\hat f^{1/2} F_{{\rm S}(\hl)}\right)'
\nn
&&
~~~=\,{1\over \hat r}F_{{\rm S}(\hm)}+
\left({\hat f'\over 2\hat f}-\hat \delta'\right)
F_{{\rm S}(\hn)}+
\hp' F_{{\rm S}(\hp)}
\,.
~~~~~
\end{eqnarray}
That is, the above four Eqs. (\ref{eq_phi}) $\sim$ (\ref{eq_mu})
are not independent.
Hence, if we  solve three of them, the remaining one equation
is automatically satisfied.

%%%%%%%%%%%%%%%%%%%%%%%%%%%%%%%%%%%%%%%%%%%%%%%%%%%%%%%%%%%%%%%%%%%
%%%%%%%%%%%%%%%%%%%%%%%%%%%%%%%%%%%%%%%%%%%%%%%%%%%%%%%%%%%%%%%%%%%
\subsection{Boundary conditions}
\label{boundary_DEGB}
%%%%%%%%%%%%%%%%%%%%%%%%%%%%%%%%%%%%%%%%%%%%%%%%%%%%%%%%%%%%%%%%%%%
%%%%%%%%%%%%%%%%%%%%%%%%%%%%%%%%%%%%%%%%%%%%%%%%%%%%%%%%%%%%%%%%%%%
In order to find a black hole solution, we need the boundary conditions
both at the event horizon and at the infinity.
Since we are interested in  an asymptotically ``flat" spacetime,
we assume
\begin{eqnarray}
\hat f&\rightarrow& 1-\left[{2\kappa_D^2
 \over (D-2)A_{D-2}}\right]{\hat M\over \hat r^{D-3}},
\nn
\hat \delta
&\rightarrow& {\cal O}\left({1\over \hat r^{D-3}}\right),
\nn
\hat \phi&\rightarrow&  {\cal O}\left({1\over \hat r^{D-3}}\right)
\,,
\end{eqnarray}
as $\hat r\rightarrow \infty$, where
\begin{eqnarray}
A_N=2\pi^{(N+1)/2}/\Gamma[(N+1)/2],
\end{eqnarray}
is the area of $N$-dimensional unit sphere, and
$\hat M$ is a gravitational mass in the string frame.

At the event horizon ($\hat r_H$),
the metric function $\hat f$ vanishes,
i.e., $\hat f(\hat r_H)= 0$.
The variables and their derivatives must be finite at $\hat r_H$.
Taking the limit of $\hat r\rightarrow \hat r_H$,
we have three independent constraints from the basic equations:
\begin{widetext}
\begin{eqnarray}
&&
\hat \rho_H^2
\left[(D-2)_3-2(D-2)\hx_H+2\hat \zeta_H+4\hx_H\hat \eta_H\right]
+\left[(D-4)_5 -4(D-4)\hx_H+4\hat \zeta_H+2\hx_H^2\right]=0,
\label{BC1}
\\
&&
\hat \rho_H^2
\left[(D-2)_3-(D-2)\hx_H+2\hx_H\hat \eta_H \right]
+ \left[(D-4)_5 -2(D-4)\hx_H+4\hx_H\hat \eta_H\right]=0,
\label{BC2}
\\
&&
\hat \rho_H^2
\left[(D-2)_4-2(D-2)_3\hx_H+2(D-2)\hat \zeta_H+4(D-2)\hx_H\hat \eta_H \right]
\nn
&&~~~~+
\left[(D-4)_6 -4(D-4)_5\hx_H+4(D-4)\hat \zeta_H
+8(D-4)\hx_H\hat \eta_H+2(D-4)\hx_H^2-8\hx^2\hat \eta_H\right]=0,
~~~~~~
\label{BC3}
\end{eqnarray}
where we have denoted the variables at the horizon with the subscript $H$,
i.e., $\hat \phi_H$, $\hat \phi_H'$,
$\hat f_H'$,$\hat \delta_H$,
$\hat \delta_H'$,
$ (\hat X\hat f)_H$ and so on,
and introduced the dimensionless variables as
\begin{eqnarray}
\hat \rho_H:=
\hat r_H/\ta_2^{1/2}~,~~
\hx_H:=\hat r_H\hat f_H'~,~~
\hat \eta_H:= \hat r_H\hat \phi_H'~,~~
\hat \zeta_H:= \hat r_H^2 (\hat X\hat f)_H
\,.
\end{eqnarray}
Eliminating $\hx_H$ and $\hat \zeta_H$ in Eqs.
(\ref{BC1}) $\sim$ (\ref{BC3})
[we assume that $\hx_H\neq 0$ and $\hat \zeta_H\neq 0$],
we find the quadratic equation for $\hat \eta_H$:
\begin{eqnarray}
\hat a\hat \eta_H^2+\hat b\hat \eta_H+\hat c=0
\label{eq_phip1}
\,,
\end{eqnarray}
where
\begin{eqnarray}
\hat a&=&4 ( \hat \rho_H^2 +2 )
\Bigl[ (D-2)_3  \hat \rho_H^6
 +2 ( D^3 -5D^2 +2D +14) \hat \rho_H^4
 +2 (D-4) (3D^2 -14D +7 ) \hat \rho_H^2 +4 (D-3)_5 \Bigr],
\nn
\hat b&=&-2 \Bigl[ (D-3) (D-2)^2 \hat \rho_H^8
+(D-2) ( 7 D^2 - 43 D + 72 )  \hat \rho_H^6
 + 2 (D-4) ( D^3 -37 D +72 )  \hat \rho_H^4
\nn
&&
 +2 (D-4) (D-5) ( 3 D^2 -5 D -16 ) \hat \rho_H^2
 +4 (D-1) (D-5) (D-4)^2 \Bigr],
\nn
\hat c&=&- (D-1) \hat \rho_H^2 \Bigl[ -(D-2)^3 \hat \rho_H^4
 +4 (D-4) (D-2) \hat \rho_H^2 +2 (D-4)^2 (D+1) \Bigr]
\,.
\end{eqnarray}
\end{widetext}

\begin{table}[ht]
\caption{The allowed values for a regular horizon radius are shown
($\hat \rho_H:= \hat r_H/\ta_2^{1/2}$).
The equality gives a double root of $\phi_H'$.
There is a minimum radius in four dimensions. In five dimensions and six dimensions, there are gaps
in which there is no regular horizon. For dimensions higher
than six, a regular horizon always exists for any horizon radius.
}
\begin{center}
\begin{tabular}{|c|l|}
\hline
$D$& ~~Condition for regular horizon
\\
\hline
\hline
4&~~$\hat \rho_H \geq 2.95712 $~~
\\
\hline
5&~~$\hat \rho_H\geq 2.55757 \,~~
{\rm or}~~~\hat \rho_H\leq  1.03572 $~~
\\
\hline
6&~~$\hat \rho_H\geq 2.25772 \,~~
{\rm or}~~~\hat \rho_H\leq  1.46781 $~~
\\
\hline
~$7\leq D\leq 10$~& ~~any values
\\
\hline
\end{tabular}
\label{table_1}
\end{center}
\end{table}

The discriminant of the quadratic Eq.~(\ref{eq_phip1})
depends on $D$ and $\hat \rho_H$.
If the discriminant is negative
($\hat {\cal D}_D:= \hat b^2-4\hat a\hat c < 0$),
 there is no real value of $\phi_H'$,
which means that no regular horizon  exists.
The condition for the discriminant to be non-negative
gives some constraint on
$\hat \rho_H^2$ for given $D$.
Since $\ta_2$ is a fundamental coupling constant,
it gives some condition on the horizon radius $\hat r_H$.

For $D=4\sim 10$, assuming $\ta_2>0$, we find
allowed values for the regular event horizon, which are summarized in Table
\ref{table_1}.
There is a minimum horizon radius $\hat r_H = 2.95712 ~\ta_2^{1/2}$
in four-dimensional spacetime,
while in five-dimensional and six-dimensional spacetimes, there is a small gap in the parameter space of
horizon radius ($1.03572 ~\ta_2^{1/2} < \hat r_H< 2.55757 ~\ta_2^{1/2} $ for five dimensions,
and $1.46781 ~\ta_2^{1/2} < \hat r_H< 2.25772 ~\ta_2^{1/2} $ for six dimensions)
where there is no regular horizon. For spacetime of dimensions higher
than six, there is a regular horizon for any horizon radius.
\\

%%%%%%%%%%%%%%%%%%%%%%%%%%%%%%%%%%%%%%%%%%%%%%%%%%%%%%%%%%%%%%%%%%
%%%%%%%%%%%%%%%%%%%%%%%%%%%%%%%%%%%%%%%%%%%%%%%%%%%%%%%%%%%%%%%%%%
%%%%%%%%%%%%%%%%%%%%%%%%%%%%%%%%%%%%%%%%%%%%%%%%%%%%%%%%%%%%%%%%%%
\section{Transformation to the Einstein frame}
\label{sec4}
%%%%%%%%%%%%%%%%%%%%%%%%%%%%%%%%%%%%%%%%%%%%%%%%%%%%%%%%%%%%%%%%%%
%%%%%%%%%%%%%%%%%%%%%%%%%%%%%%%%%%%%%%%%%%%%%%%%%%%%%%%%%%%%%%%%%%
%%%%%%%%%%%%%%%%%%%%%%%%%%%%%%%%%%%%%%%%%%%%%%%%%%%%%%%%%%%%%%%%%%

Here we give the relation between the variables in the string
frame and those in the Einstein frame.
The metrics in both frames are given by
Eq. (\ref{metric_string2}) and
\begin{eqnarray}
ds_D^2=-f e^{-2\delta}dt^2+{1\over f}dr^2
+r^2 d\Omega_{D-2}^2
\label{metric_Einstein2}
\,,
\end{eqnarray}
respectively.
The conformal transformation (\ref{conf_trans}) or
\begin{eqnarray}
\hat  g_{\mu\nu}&=&\exp[ \gamma \phi]\, g_{\mu\nu},
\label{conf_trans2}
\end{eqnarray}
gives the relation between the metric components as follows:
\begin{eqnarray}
\hat r&=&\exp \left[\gamma\phi \over 2\right] \, r ,
\label{radial1}
\\
\hat f&=&\left(1+{\gamma r\over 2}{ d\phi\over dr}\right)^2 \,f ,
\label{potential1}
\\
\hat \delta&=&\delta
-{\gamma\phi\over 2}
+\ln \left(1+{\gamma r\over 2}{d\phi\over dr}\right)
\label{lapse1}
\,,
\end{eqnarray}
or inversely,
\begin{eqnarray}
r&=&\exp [-\gamma^2\hat \phi] \, \hat r ,
\label{radial2}
\\
f&=&\left(1-\gamma^2 \hat r {d\hat \phi\over d\hat r}\right)^2
\,\hat f ,
\label{potential2}
\\
\delta&=&\hat \delta
+\gamma^2 \hat \phi+\ln \left(1-\gamma^2
\hat r{d\hat \phi\over d\hat r}\right)
\label{lapse2}
\,.
\end{eqnarray}
Since the radial coordinates $r$ and $\hat r$ are related
by Eq. (\ref{radial1}) or Eq. (\ref{radial2}),
the horizon radii must be rescaled:
\begin{eqnarray}
r_H&=&\exp [-\gamma^2\hat \phi_H] \, \hat r_H ,
\nn
\hat r_H&=&\exp \left[\gamma\phi_H\over 2\right] \, r_H .
\end{eqnarray}

The gravitational masses and
the scalar charges are also rescaled as
\begin{eqnarray}
GM&=&G\hat M-(D-3)\c^2\hat Q_\phi ,
\nn
G\hat M&=&GM+{(D-3)\c\over 2} Q_\phi,
\\
 Q_\phi
&=&2\c \hat Q_\phi
\,,
\end{eqnarray}
where the scalar charges $\hat Q_\phi$ in the string frame
and $Q_\phi$ in
 the Einstein frame are defined by the asymptotic behaviours
\begin{eqnarray}
\hat \phi&\rightarrow&\left[{8\pi
 \over (D-2)A_{D-2}}\right]
{\hat Q_\phi\over \hat r^{D-3}},
\nn
\phi&\rightarrow&\left[{8\pi
 \over (D-2)A_{D-2}}\right]
{Q_\phi\over r^{D-3}}
\,,
\end{eqnarray}
as $\hat r \rightarrow \infty\, (r\rightarrow \infty)$, respectively.

The lapse function $\delta$ in the Einstein frame
must drop as $1/r^{2(D-3)}$. As a result, we find
\begin{eqnarray}
\hat \delta&\rightarrow&-
\left[{8\pi\c \over A_{D-2}}\right]
\frac{\hat Q_\phi}{ \hat r^{D-3}},
\nn
\delta&\rightarrow&{\cal O}\left(\frac{1}{ r^{2(D-3)}}\right)
\,,
\end{eqnarray}
near infinity.

%%%%%%%%%%%%%%%%%%%%%%%%%%%%%%%%%%%%%%%%%%%%%%%%%%%%%%%%%%%%%%%%%%
%%%%%%%%%%%%%%%%%%%%%%%%%%%%%%%%%%%%%%%%%%%%%%%%%%%%%%%%%%%%%%%%%%
%%%%%%%%%%%%%%%%%%%%%%%%%%%%%%%%%%%%%%%%%%%%%%%%%%%%%%%%%%%%%%%%%%
\section{Thermodynamic Variables}
\label{sec5}
%%%%%%%%%%%%%%%%%%%%%%%%%%%%%%%%%%%%%%%%%%%%%%%%%%%%%%%%%%%%%%%%%%
%%%%%%%%%%%%%%%%%%%%%%%%%%%%%%%%%%%%%%%%%%%%%%%%%%%%%%%%%%%%%%%%%%
%%%%%%%%%%%%%%%%%%%%%%%%%%%%%%%%%%%%%%%%%%%%%%%%%%%%%%%%%%%%%%%%%%

Before showing our numerical result, let us
introduce thermodynamical variables of a black hole.

The Hawking temperature is given from
 the periodicity of the Euclidean time
on the horizon as
\begin{align}
\hat T_{\rm H} &= \frac{1}{4\pi}\hat f'_He^{-\hat \d_H}
\,,
\nn
T_{\rm H} &= \frac{1}{4\pi}f'_He^{-\d_H}
\,.
\end{align}
Although we can define the Hawking temperature both in
the string frame and in the Einstein frame, we find that
both temperatures are the same by using the relation
\begin{eqnarray}
\hat f_H'=\left(1+\frac{\c r_H }{ 2}\phi_H'\right)e^{-\frac{\c \phi_H}{ 2}}f_H'
\,.
\end{eqnarray}
As for a black hole entropy, it does not obey the Bekenstein-Hawking formula, i.e.,
a quarter of the area of event horizon, because
we have the Gauss-Bonnet terms.
According to the Wald's formula for a black hole entropy,
which is defined by use of
the Noether charge associated with the diffeomorphism
invariance of the system~\cite{Wald_formula}, we find
\begin{align}
S = -2 \pi \int_\Sigma \frac{\pa {\cal L}}{\pa R_{\mn \rho\s}}
\e_{\mn} \e_{\rho\s},
\end{align}
where $\Sigma$ is $(D-2)$-dimensional surface of the event horizon,
${\cal L}$ is the Lagrangian density,
$\e_\mn$ denotes the volume element binormal to $\Sigma$.

For the effective action in the string frame (\ref{Saction}), it gives
\begin{align}
S_S &= \frac{e^{-2 \hp_H} \hat{A}_H }{4} \lh 1 +\frac{2\ta_2}{\hr_H^2}
\rh,
\end{align}
where $\hat A_H$ is the area of the event horizon.
Using the variables in the Einstein frame, this entropy is
rewritten as~\cite{ohta_torii1}
\begin{align}
S_{\rm S} = \frac{A_H}{4} \lh 1 +\frac{2\ta_2}{r_H^2}
e^{-\c \phi_H} \rh.
\end{align}

We may look at the corrections to the
Bekenstein-Hawking entropy ($S_{\rm BH}:=A_H/4$) which is
\begin{align}
&&
S_{\rm S}-S_{\rm BH}
=\frac{2\ta_2 }{r_H^2}e^{-\c \phi_H}\times S_{\rm BH}
>0
\label{entropy_string}
\,.
\end{align}
$S_{\rm S}$ is always larger than $S_{\rm BH}$.

%%%%%%%%%%%%%%%%%%%%%%%%%%%%%%%%%%%%%%%%%%%%%%%%%%%%%%%%%%%%%%%%%%
%%%%%%%%%%%%%%%%%%%%%%%%%%%%%%%%%%%%%%%%%%%%%%%%%%%%%%%%%%%%%%%%%%
\section{Black Holes in The DEGB theory}
\label{sec6}
%%%%%%%%%%%%%%%%%%%%%%%%%%%%%%%%%%%%%%%%%%%%%%%%%%%%%%%%%%%%%%%%%%
%%%%%%%%%%%%%%%%%%%%%%%%%%%%%%%%%%%%%%%%%%%%%%%%%%%%%%%%%%%%%%%%%%
Now we present our numerical results.

Giving a horizon radius $r_H$, we solve the basic equations.
To solve the equations numerically,
we first set $\delta_H=0,\phi_H=0$, and find the asymptotically
flat spacetime. We then rescale
the lapse function and the dilaton field
as $\tilde \delta(r)=\delta(r)-\delta(\infty)$
and
$\tilde \phi(r)=\phi(r)-\phi(\infty)$.
This is always possible because $\delta$ and $\phi$ appear only in the
forms of their derivatives such as $\delta'$ and $\phi'$.
As a result, we can set $\tilde \delta(r), \tilde \phi(r) \rightarrow 0$
for $r\rightarrow \infty$.
Then in our actual solutions, $\delta_H$ and $\phi_H$ do not vanish.
In what follows, for brevity, we omit the tilde of the variables.

Depending on the dimension,
we classify our solutions into three types:
(a) four dimensions ($D=4$),
(b) five dimensions ($D=5$), and
(c) six or higher dimensions ($D=6 \sim 10$).

%%%%%%%%%%%%%%%%%%%%%%%%%%%%%%%%%%%%%%%%%%%%%%%%%%%%%%%%%%%%%%%%%%
%%%%%%%%%%%%%%%%%%%%%%%%%%%%%%%%%%%%%%%%%%%%%%%%%%%%%%%%%%%%%%%%%%
\subsubsection{Mass-area relation}
%%%%%%%%%%%%%%%%%%%%%%%%%%%%%%%%%%%%%%%%%%%%%%%%%%%%%%%%%%%%%%%%%%
%%%%%%%%%%%%%%%%%%%%%%%%%%%%%%%%%%%%%%%%%%%%%%%%%%%%%%%%%%%%%%%%%%

First we show the relation between the black hole mass $M$ and
the horizon area $A_H$ in Fig.~\ref{mass-area_DEGB}.
When we give the numerical result, we show $\bar M:=\kappa_D^2 M$
for the mass instead of the gravitational mass $M$,
because its value can be scaled as $\tilde \alpha_2^{(D-3)/2}$
when we change the coupling constant $\alpha_2$.
In the unit of $\kappa_D=1$, both masses are equivalent.
Hence, in what follows, for simplicity, we do not distinguish two masses
and use $M$ for both masses.

In the four-dimensional case, as shown in Fig.\ref{mass-area_DEGB} (a),
there is the minimum radius below which there is no black hole
($\hat r_{H (\rm min)}=2.95712\ta_2^{1/2}$).
The ranges of the horizon radius where the black holes exist are
shown in Table \ref{range_regular_BH} and are narrower than
those from the regularity condition in Table~\ref{table_1} in general,
%This is expected from the regularity condition of the horizon.
though the minimum radius in the string frame coincides
with the value in Table~\ref{table_1} for four dimensions.
The minimum mass of the black hole shown in
Fig.\ref{mass-area_DEGB}(a) is given by $M_{\rm min}=69.3511\ta_2^{1/2}$.
Near this minimum mass, we find that the $M$-$A_H$ curve turns around,
i.e., there are two black hole solutions for a given mass
($M_{\rm min}\leq M<72.3945\ta_2^{1/2}$).
We suspect that the larger black hole is stable, while the smaller one is unstable
(see the later discussion about the entropy).
\begin{widetext}
\begin{center}
\begin{figure}[ht]
\includegraphics[width=58mm]{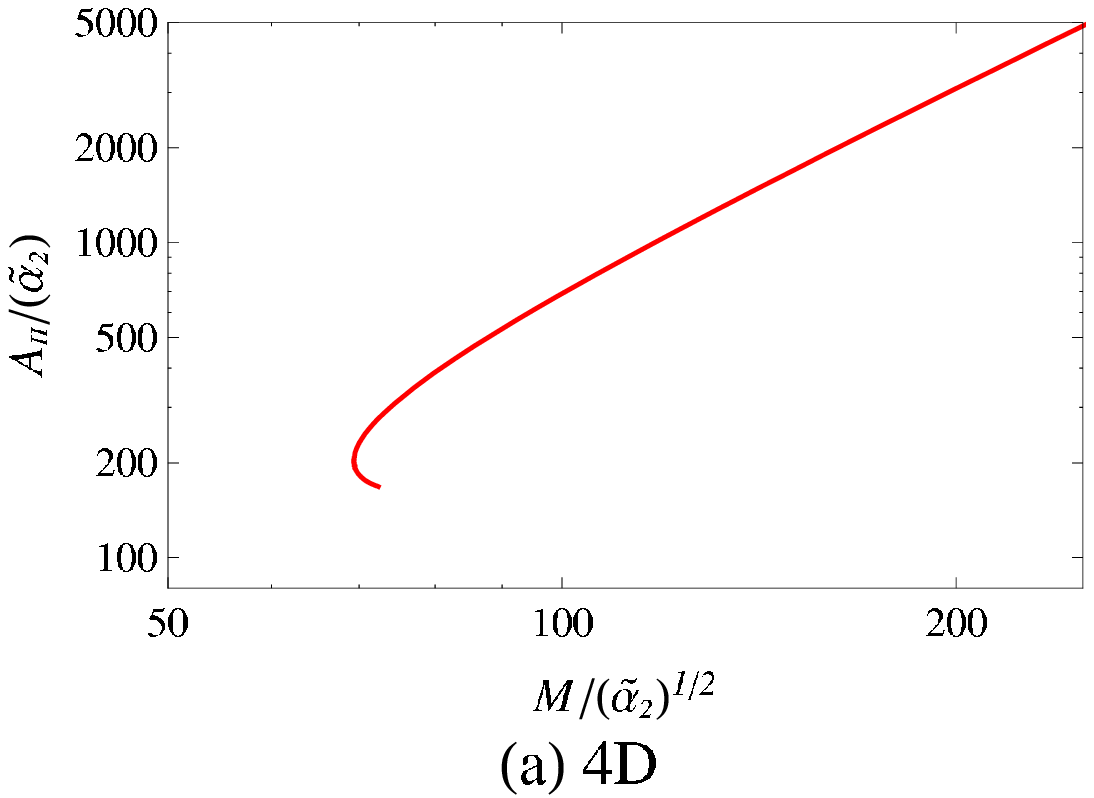}~~~
\includegraphics[width=58mm]{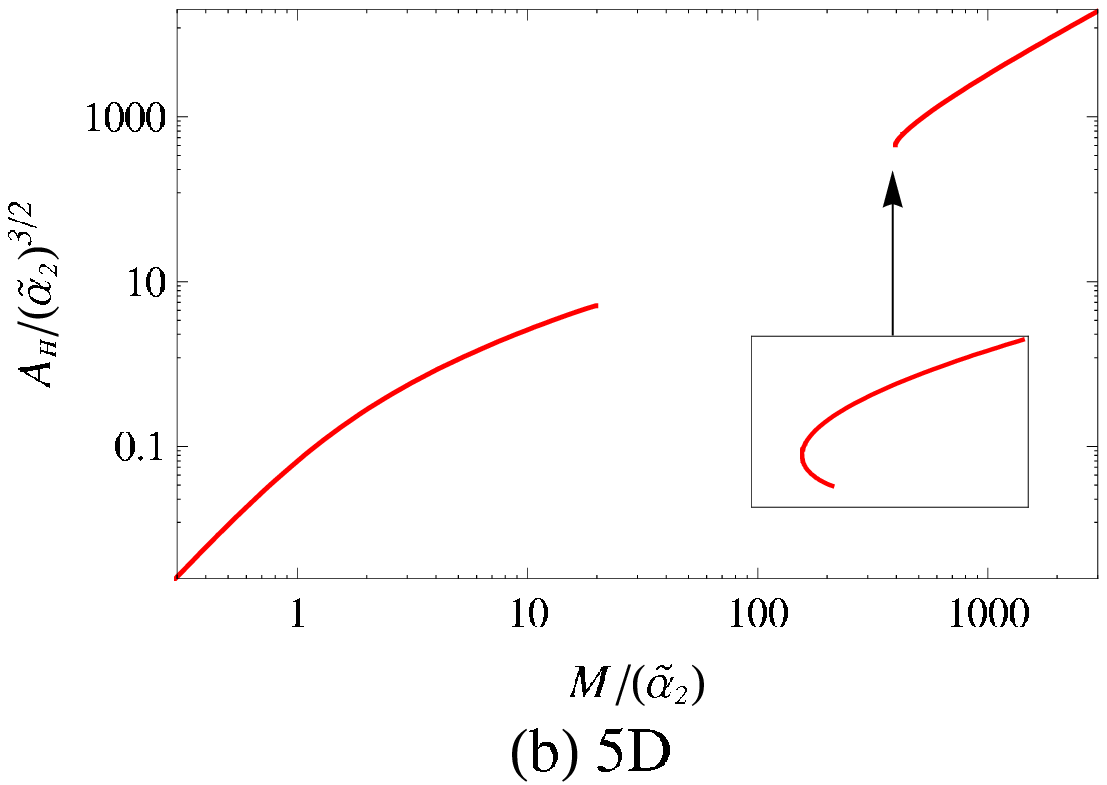}~~~
\includegraphics[width=58mm]{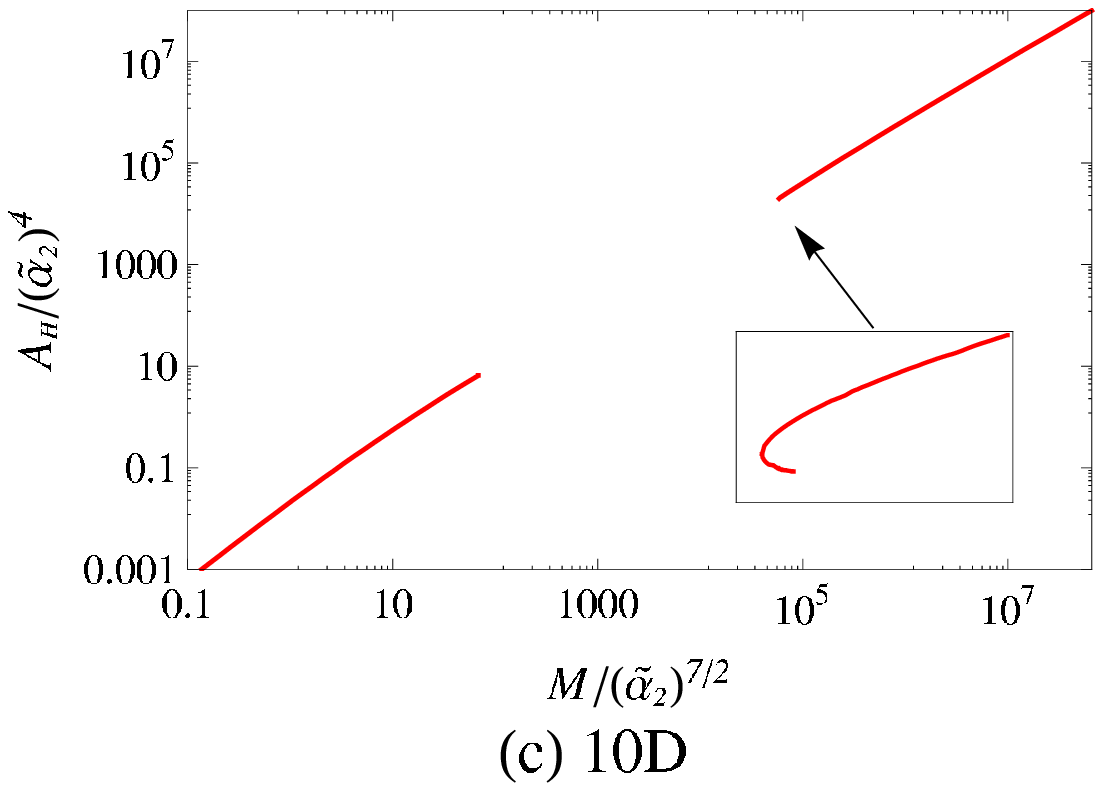}
\caption{The horizon area $A_H$ in terms of the gravitational mass $M$.
There appear mass gaps in five dimensions and ten dimensions.
In four dimensions and in the L-branches of five dimensions and ten dimensions,
we find the $M$-$A_H$ curves turn-around the minimum masses.
We have two black hole solutions near the turn-around points.
There is no turn-around behaviour in the S-branch.
}
\label{mass-area_DEGB}
\end{figure}
\end{center}
\vskip -1cm
\begin{table}[ht]
\caption{The ranges of the horizon radii in which numerically
solved black hole solutions exist both in the string frame
($\hat \rho_H:= \hat r_H/\ta_2^{1/2}$)
and in the Einstein frame ($\rho_H:= r_H/\ta_2^{1/2}$).
They are related by the conformal factor.
There is a minimum radius in four dimensions. 
For dimensions higher than four dimensions,
there is a gap, in which there is no regular black hole solution.
The ranges of the horizon radius in the string frame
should be compared with those in Table~\ref{table_1}
}
\begin{center}
\begin{tabular}{|c|l|l|}
\hline
$D$& ~~~~~~~~~~~~~~~~~~~~~String frame
&~~~~~~~~~~~~~~~~~Einstein frame
\\
\hline
\hline
4&~~$\hat \rho_H\geq {2.95712}  $~~
&~~$\rho_H\geq {3.65726} $~~
\\
\hline
5&~~$\hat \rho_H\geq {2.55757} \,~~
{\rm or}~~~\hat \rho_H\leq {0.647144} $~~
&~~$\rho_H\geq {2.84836} \,~~
{\rm or}~~~ \rho_H\leq {0.636670} $~~
\\
\hline
6&~~$\hat \rho_H\geq {2.26141} \,~~
{\rm or}~~~\hat \rho_H\leq {0.780228} $~~
&~~$\rho_H\geq {2.44467} \,~~
{\rm or}~~~ \rho_H\leq {0.788059} $~~
\\
\hline
7&~~$\hat \rho_H\geq {2.13495} \,~~
{\rm or}~~~\hat \rho_H\leq {0.835000} $~~
&~~$\rho_H\geq {2.25907} \,~~
{\rm or}~~~ \rho_H\leq {0.842271} $~~
\\
\hline
8&~~$\hat \rho_H\geq {2.12365} \,~~
{\rm or}~~~\hat \rho_H\leq {0.848702} $~~
&~~$\rho_H\geq {2.21291} \,~~
{\rm or}~~~ \rho_H\leq {0.853517} $~~
\\
\hline
9&~~$\hat \rho_H\geq {2.15406} \,~~
{\rm or}~~~\hat \rho_H\leq {0.842414} $~~
&~~$\rho_H\geq {2.22300} \,~~
{\rm or}~~~ \rho_H\leq {0.845275} $~~
\\
\hline
10&~~$\hat \rho_H\geq {2.20216} \,~~
{\rm or}~~~\hat \rho_H\leq {0.827220} $~~
&~~$\rho_H\geq {2.25818} \,~~
{\rm or}~~~ \rho_H\leq {0.828790} $~~
\\
\hline
\end{tabular}
\label{range_regular_BH}
\end{center}
\end{table}
\end{widetext}

In the five-dimensional case, there appears new type of solutions near the zero-mass
region as shown in Fig. \ref{mass-area_DEGB}(b). We find two mass ranges:
one has the smaller masses (S-branch ),
and the other has the larger masses (L-branch).
There is a mass gap between these two branches.
It has been expected from the results which
we found from the regularity condition for the horizon (see Table \ref{table_1}).
The L-branch is similar to the solutions in the four-dimensional case.
There exists the lower mass bound.
Near the minimum mass ($M_{\rm min}^{\rm (L)}=395.862\ta_2$),
we find two black hole solutions
in the range of $M_{\rm min}^{\rm (L)}<M<395.880\ta_2$
(see the enlarged figure in Fig. \ref{mass-area_DEGB}(b)).
The minimum radius in this branch is found by the existence
condition of the regular horizon, i.e.,
$\hat r_{H (\rm min)}=2.55757\ta_2^{1/2}$ (Compare Tables \ref{table_1}
and \ref{range_regular_BH}) just as in the four-dimensional case.
In the S-branch, we find the maximum mass
($M_{\rm max}^{\rm (S)}=19.7733\ta_2$).
There is no turn-around behaviour near the maximum mass in the S-branch.
As the horizon radius approaches zero, the gravitational mass vanishes.
We find $M\approx 0.100295 \,
\ta_2^{1/2} r_H$ in the zero-mass limit.
(Note that $M\propto r_H^{2}$ in the case of
five-dimensional Schwarzschild black hole.)

In dimensions higher than five, we find the similar
structures; i.e., there are two branches (the S- and L-branches).
However, in the L-branch, the minimum radius is not given
 by the regularity condition (see Tables \ref{table_1}
and \ref{range_regular_BH}).
In fact, for $D\geq 7$, we find a gap in the range of black hole radii
in numerical solutions, but the regular horizon
condition is always satisfied for
any horizon radii. In this gap, we cannot find any
asymptotically flat black hole solution,
although the horizon can be regular.
The L-branch shows the similar properties to those in
the four-dimensional or five-dimensional case.
In the S-branch in ten dimensions, we find
$M_{\rm max}^{\rm (S)}=68.6614\ta_2^{7/2}$,
 and $M\approx 84.1890\,\ta_2 r_H^{5}$ in the zero-mass limit.
(Note that $M\propto r_H^{7}$ in the case of
ten-dimensional Schwarzschild black hole.)

We cannot make definite statement about what happens in the region of a mass gap.
Since there is no static black hole, the spacetime may be
always dynamical losing the mass energy
and eventually reaching the S-branch, or
it may evolve into a naked singularity.

%%%%%%%%%%%%%%%%%%%%%%%%%%%%%%%%%%%%%%%%%%%%%%%%%%%%%%%%%%%%%%%%%%
%%%%%%%%%%%%%%%%%%%%%%%%%%%%%%%%%%%%%%%%%%%%%%%%%%%%%%%%%%%%%%%%%%
\subsubsection{Thermodynamics}
%%%%%%%%%%%%%%%%%%%%%%%%%%%%%%%%%%%%%%%%%%%%%%%%%%%%%%%%%%%%%%%%%%
%%%%%%%%%%%%%%%%%%%%%%%%%%%%%%%%%%%%%%%%%%%%%%%%%%%%%%%%%%%%%%%%%%
Next we present the thermodynamical variables.
First we give the entropy in terms of the gravitational mass $M$ in
Fig.\ref{fig_entropy_DEGB}.
The entropy behaves very similar to
the area of the horizon, although there is a correction to the
Bekenstein-Hawking entropy, which is also shown
by the (blue) dotted line as a reference in the figure.
In particular, we find turn-around behaviours near the minimum masses
in four dimensions and in the L-branches of five dimensions and ten dimensions.
Near the minimum points, we have two black holes for a given mass.
The larger black hole has the larger entropy, and then we expect that
it is dynamically stable. On the other hand,
the smaller black hole has the smaller entropy, and then
we expect that it is dynamically unstable.

We also show the temperatures of the black holes in Fig.~\ref{temperature_DEG}.
The temperature in four dimensions is always finite and
shows the turn-around behaviour near the minimum mass.
At this turning point, we expect a change of stability (see
the discussion in~\cite{Katz}).
When the black hole evaporates via the Hawking radiation,
the mass decreases. Although the temperature is finite,
it does not vanish at the  minimum mass,
and the evaporation never stops at the minimum mass.
We may find a naked singularity.

In five dimensions and higher dimensions, we find the same behaviour as for the
L-branch.
In  the S-branch, however, the temperature in five dimensions is always finite
and vanishes at the zero-mass  limit.
Then the black hole may disappear via the Hawking radiation.

On the other hand, the behaviour is very different in dimensions higher than five.
The temperature in ten dimensions diverges
as $M\rightarrow 0$.
We find the same behaviour for the case of $D=6\sim 9$.
The evaporation never stops even near the zero-mass limit.
Rather the black hole may explode via the Hawking radiation.

As a result, we can classify our solutions into three types:
(a) $D=4$, (b) $D=5$, and (c) $D=6\sim 10$.

The reason why we have three types may be understood as follows:
The Gauss-Bonnet curvature in four dimensions becomes a total divergence if there is no
dilaton field, and then it does not give any contribution in the basic equations.
Even if we include a dilaton field, we expect the dynamical properties of
the Gauss-Bonnet term in four dimensions is very much different from
those in the case of $D\geq 5$, in which the Gauss-Bonnet term gives
a significant change in the basic equations without a dilaton field.

\begin{widetext}

\begin{figure}[ht]
\includegraphics[width=58mm]{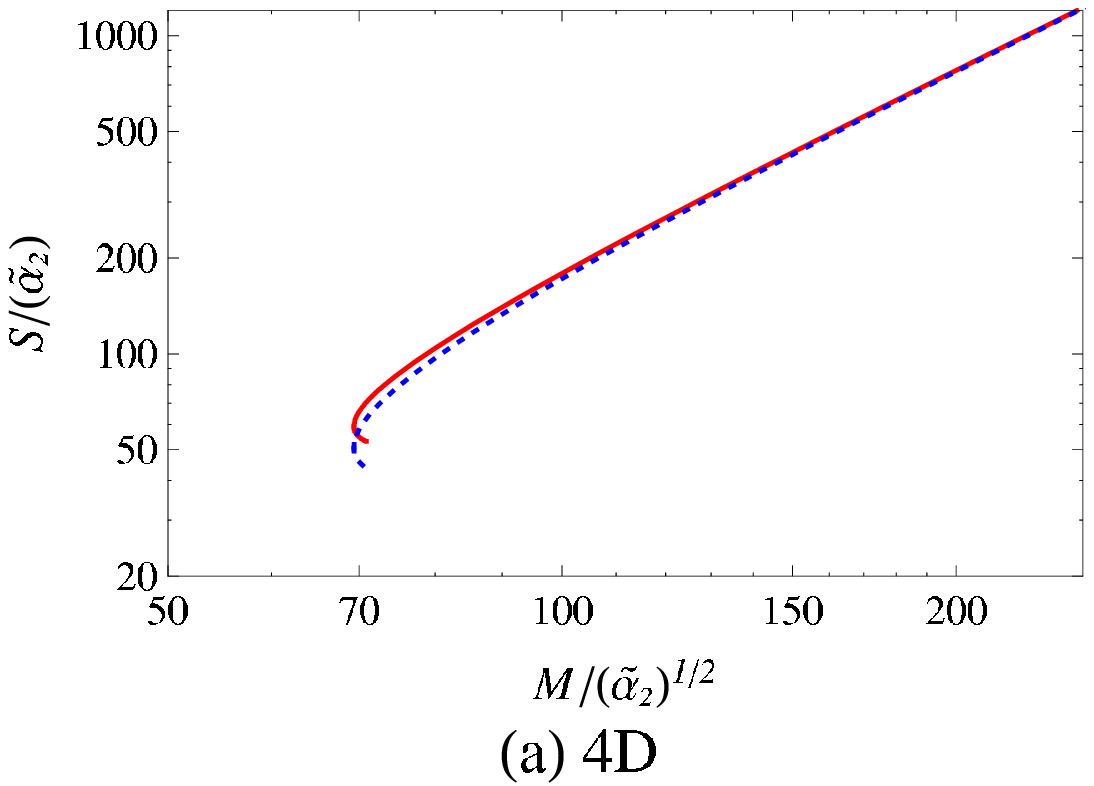}~~~
\includegraphics[width=58mm]{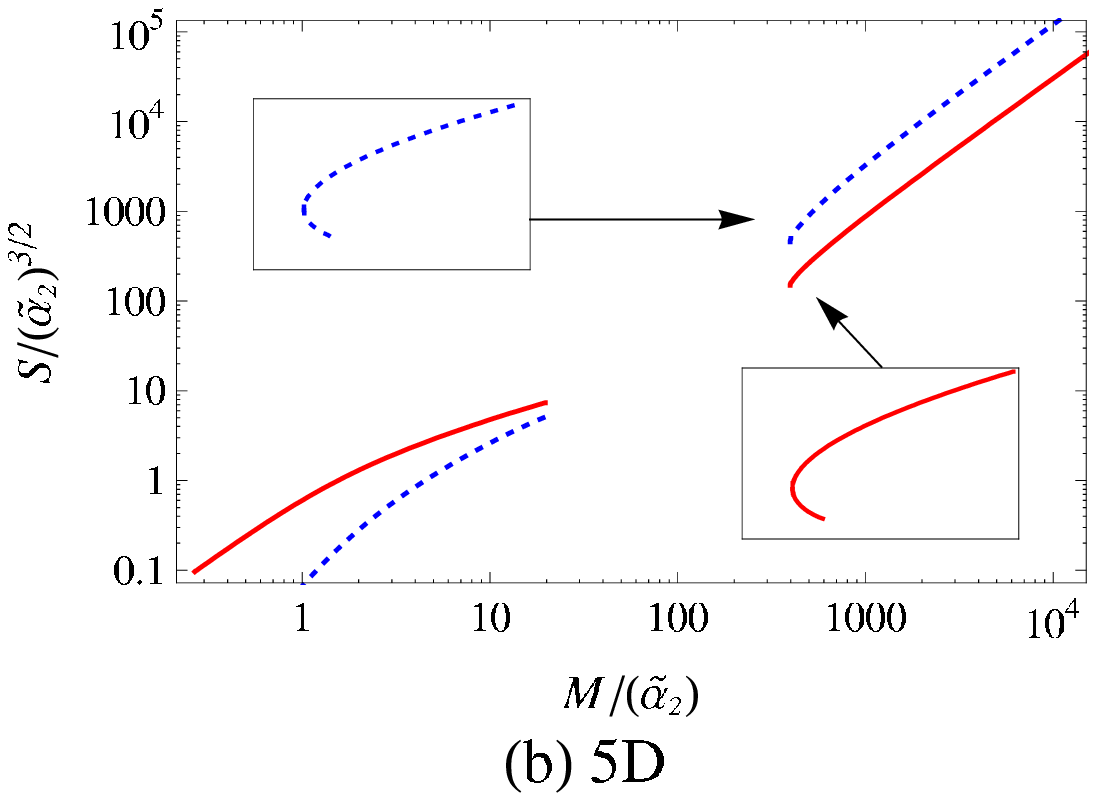}~~~
\includegraphics[width=58mm]{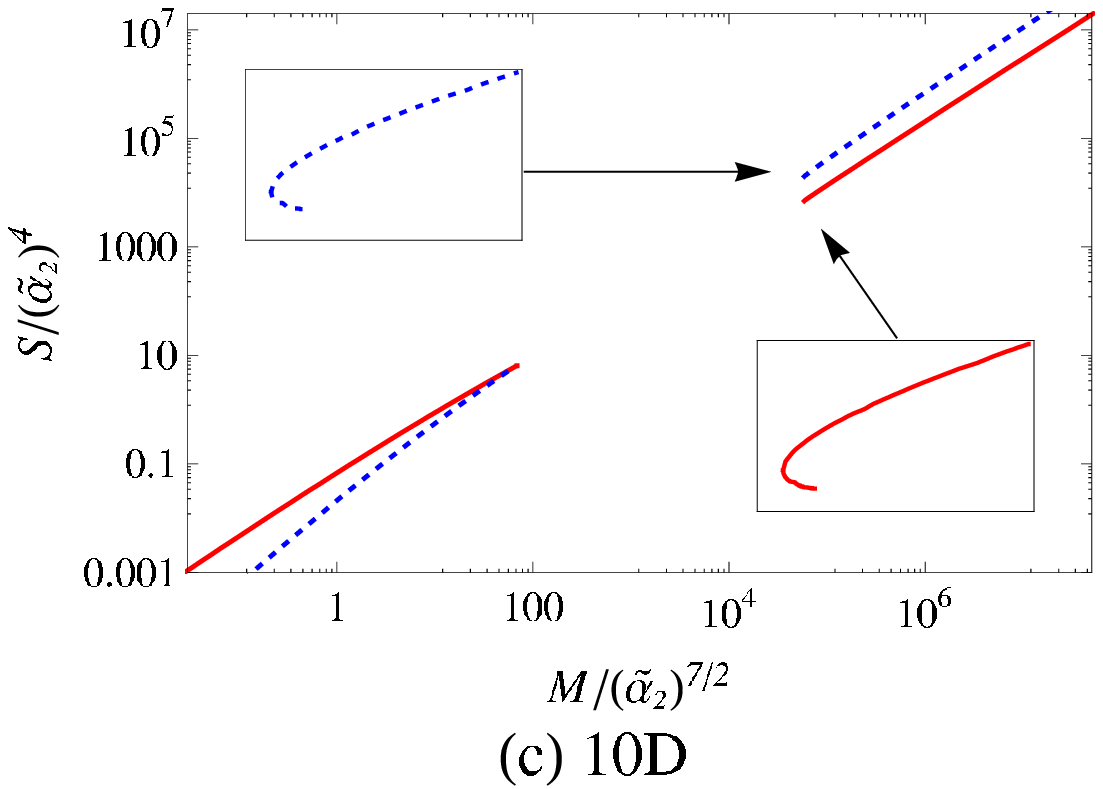}
\caption{
The entropies $S_S$  of black holes in DEGB
in terms of the mass $M$ by the solid (red) line
for (a) $D=4$, (b) $D=5$, and (c) $D=10$.
As a reference, we also show
the Bekenstein-Hawking
entropy $S_{\rm BH}=A_H/4$
by the (blue) dotted line.}
\label{fig_entropy_DEGB}
\end{figure}

\begin{figure}[ht]
\includegraphics[width=58mm]{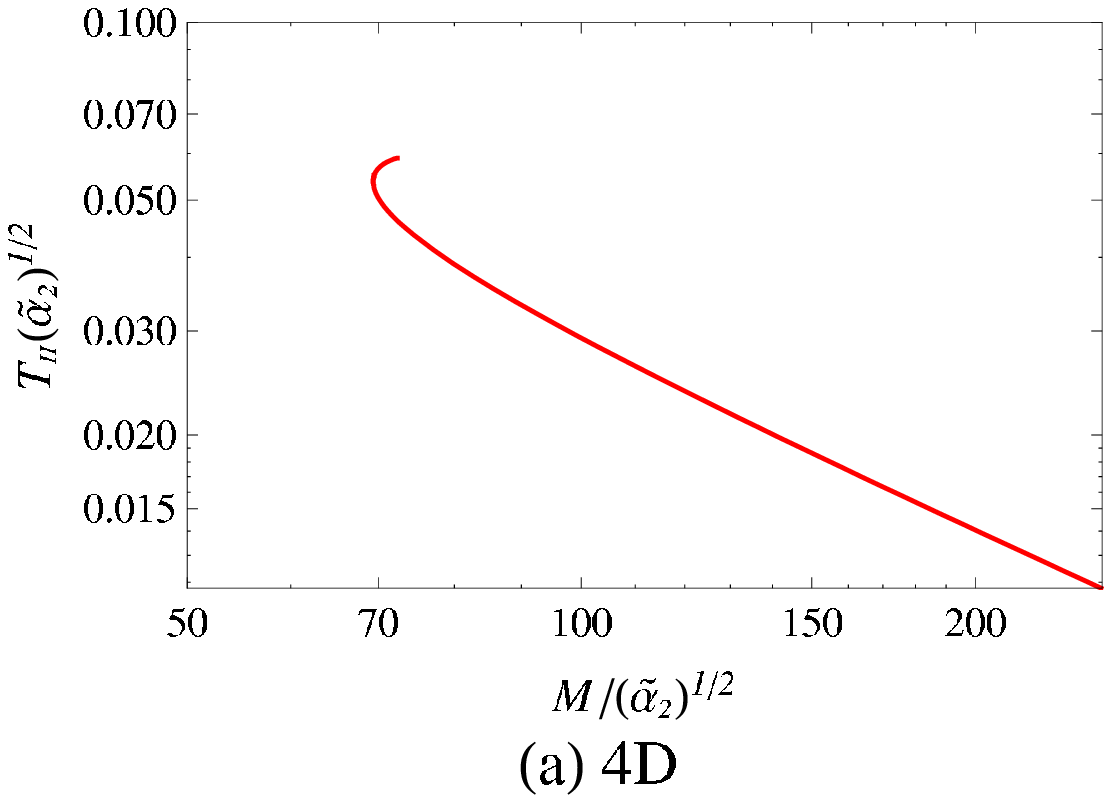}~~~
\includegraphics[width=58mm]{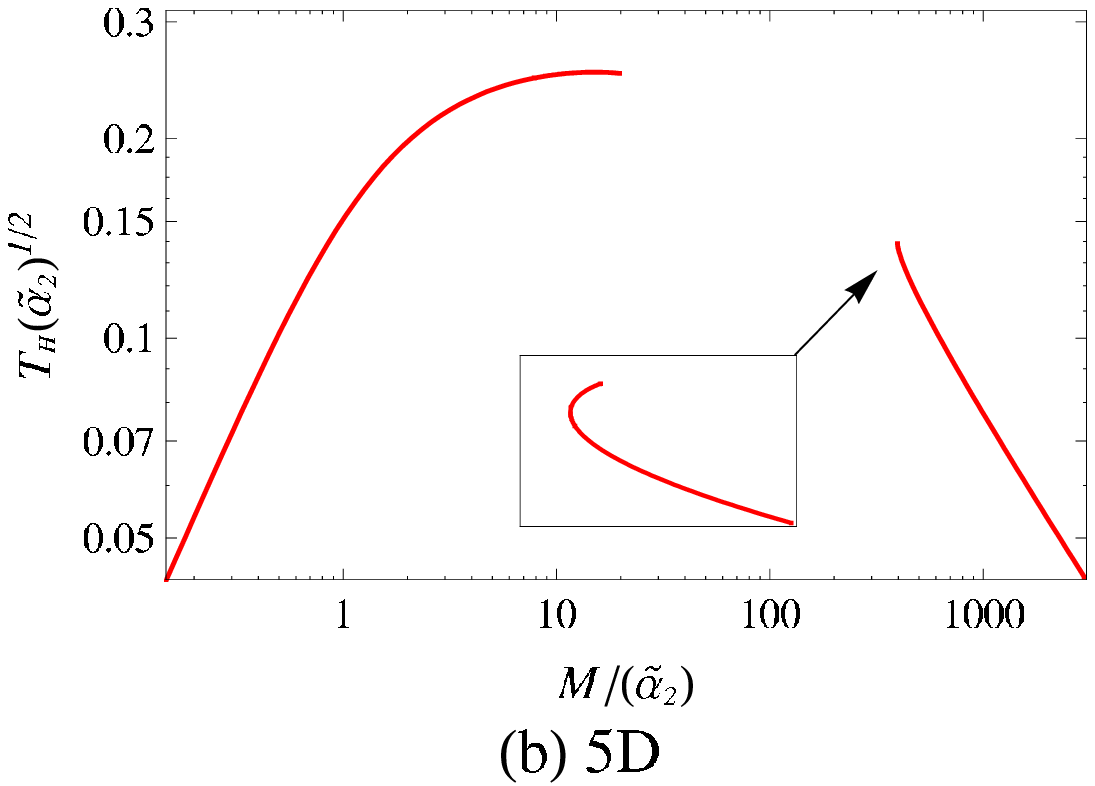}~~~
\includegraphics[width=58mm]{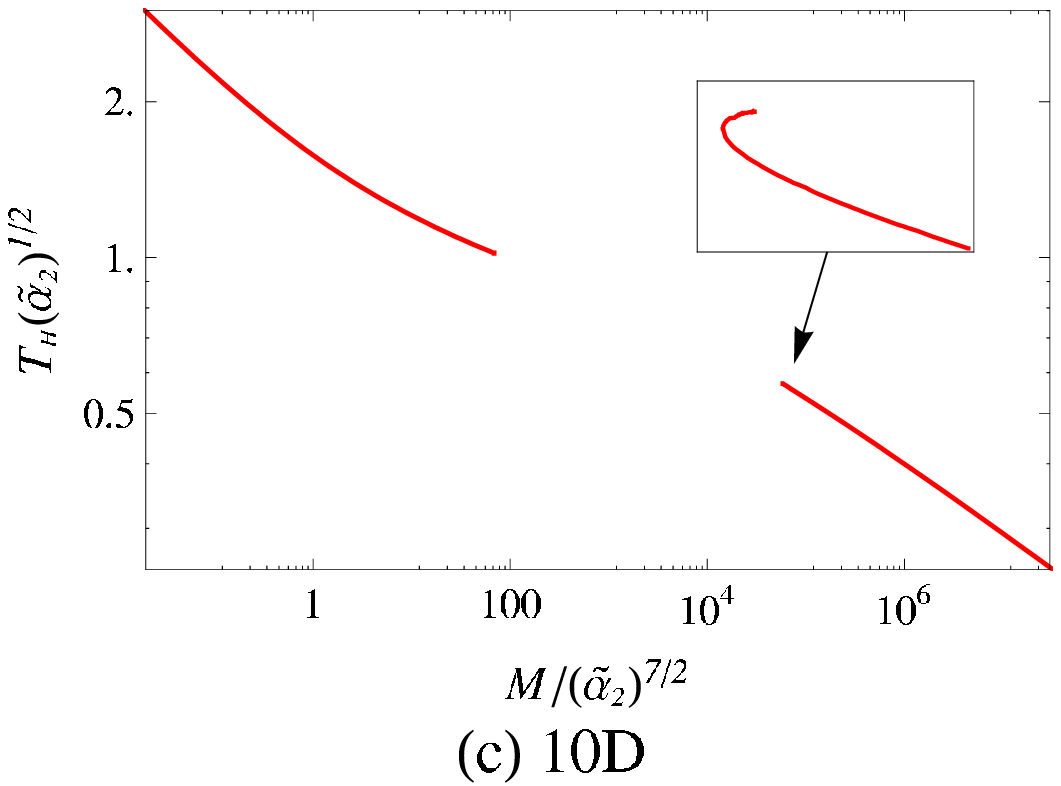}
\caption{Temperatures of black holes in DEGB
for (a) $D=4$, (b) $D=5$, and (c) $D=10$.
}
\label{temperature_DEG}
\end{figure}

Five dimensions and six dimensions are also different from other higher dimensions,
because the Gauss-Bonnet term is the highest Lovelock term in five dimensions and six dimensions,
while in higher dimensions ($D=7\sim 10$), there exist higher Lovelock terms.
There may also be a big difference between odd and even dimensions.
Hence, we expect four types: four dimensions, five dimensions and six dimensions, and higher dimensions
($D=7\sim 10$). However, it turns out that the solutions in six dimensions
and higher dimensions look similar.
As a result, we find three types.

%%%%%%%%%%%%%%%%%%%%%%%%%%%%%%%%%%%%%%%%%%%%%%%%%%%%%%%%%%%%%%%%%
%%%%%%%%%%%%%%%%%%%%%%%%%%%%%%%%%%%%%%%%%%%%%%%%%%%%%%%%%%%%%%%%%%
\subsubsection{Configuration of the metric and dilaton field}
%%%%%%%%%%%%%%%%%%%%%%%%%%%%%%%%%%%%%%%%%%%%%%%%%%%%%%%%%%%%%%%%%%
%%%%%%%%%%%%%%%%%%%%%%%%%%%%%%%%%%%%%%%%%%%%%%%%%%%%%%%%%%%%%%%%%%

Here, we show the behaviour of the mass function defined by
$m(r)=r^{D-3}( 1 - f(r) )/2
\,,
$ which approaches the ADM(Arnowitt-Deser-Miser) mass at infinity,
the lapse function $\d(r)$ and the dilaton field $\phi(r)$,
for several values of the horizon radii in Fig. \ref{4Dproperties}
 in four dimensions.

%--figures-------------------------------------------------------------
\begin{figure}[ht]
\includegraphics[width=58mm]{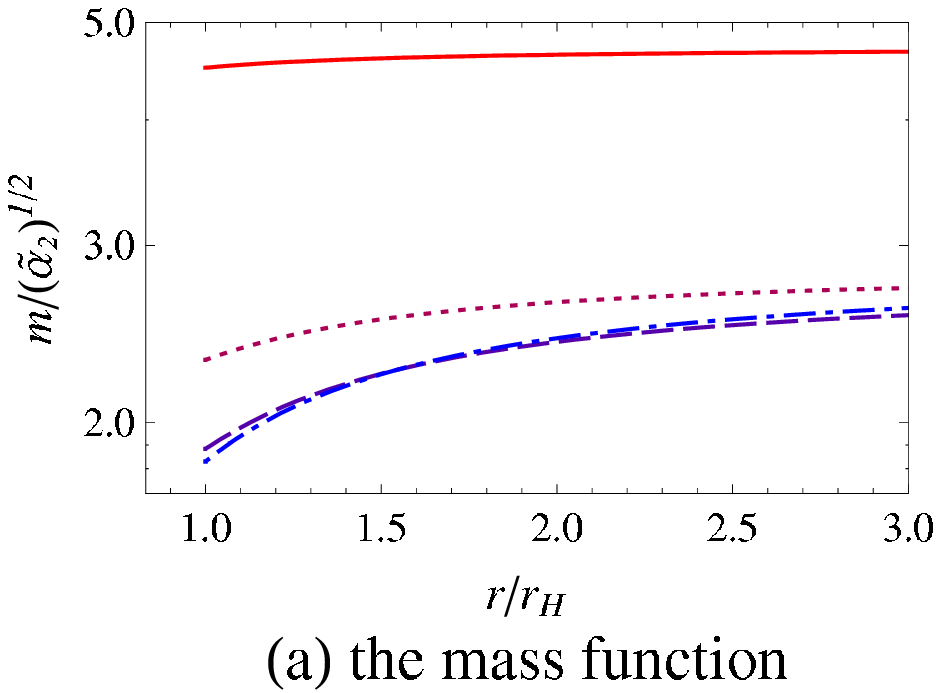}~~~
\includegraphics[width=58mm]{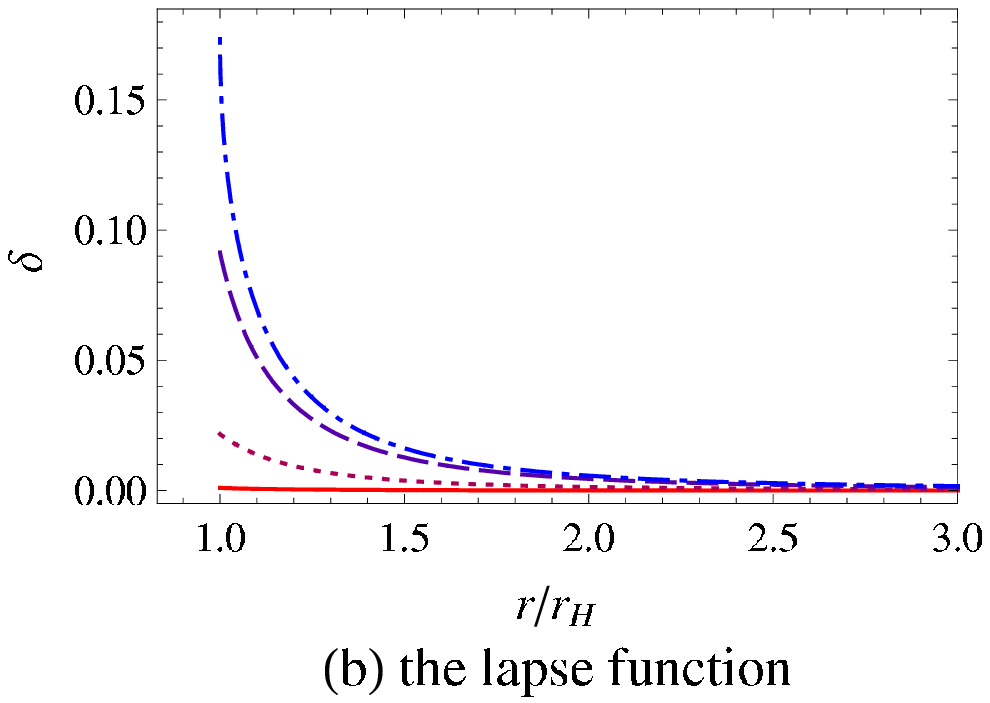}~~~
\includegraphics[width=58mm]{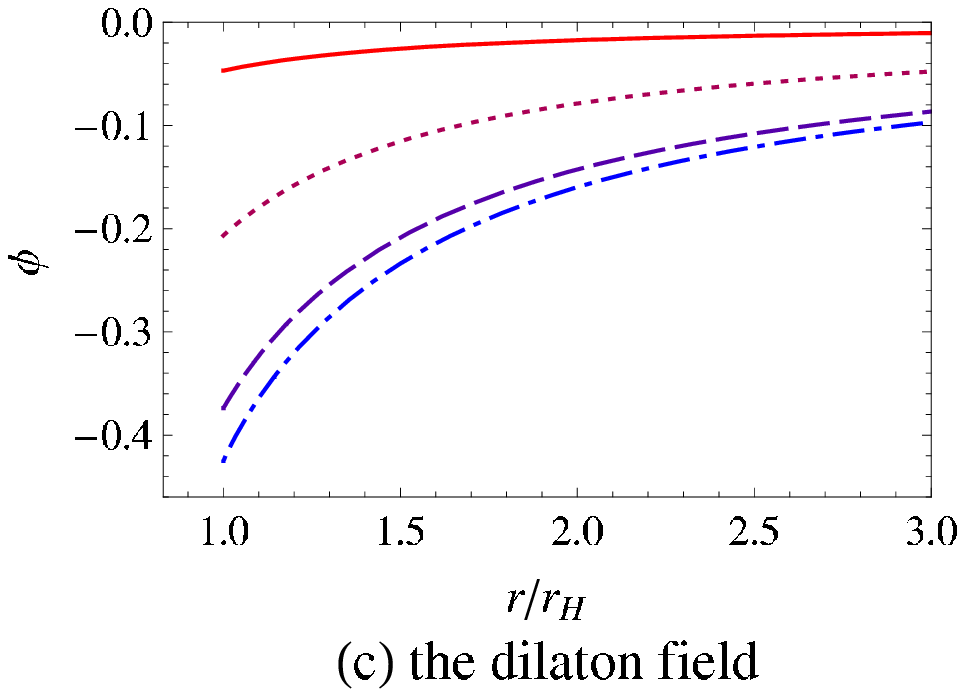}
\caption{
The solution of the four-dimensional black hole.
We choose the following four values for the horizon radius:
 $r_H=$8.80392$\ta_2^{1/2}$ (solid), 4.16383$\ta_2^{1/2}$ (dotted),
 3.12416$\ta_2^{1/2}$ (dashed), and 2.95712$\ta_2^{1/2}$ (dot-dashed).
}
\label{4Dproperties}
\end{figure}

%--figures-------------------------------------------------------------

In four dimensions, we find that
the mass function, lapse function and dilaton field behave monotonously
smooth ( see Fig. \ref{4Dproperties}).
For the solution with minimum horizon radius, however,
 the second derivative of the dilaton field  diverges
(Fig. \ref{4Dproperties} (c) ).
We show the Kretschmann curvature invariant defined by
$
K:=R_{\mu\nu\rho\sigma}R^{\mu\nu\rho\sigma}
$
for several radii in
Fig. \ref{5D_singularity} (a).
We can see the curvature at the horizon
increases rapidly near the minimum radius.
We find the curvature singularity on the horizon
in the limit of $M\rightarrow M_{\rm min}$.
\begin{figure}[ht]
\includegraphics[width=58mm]{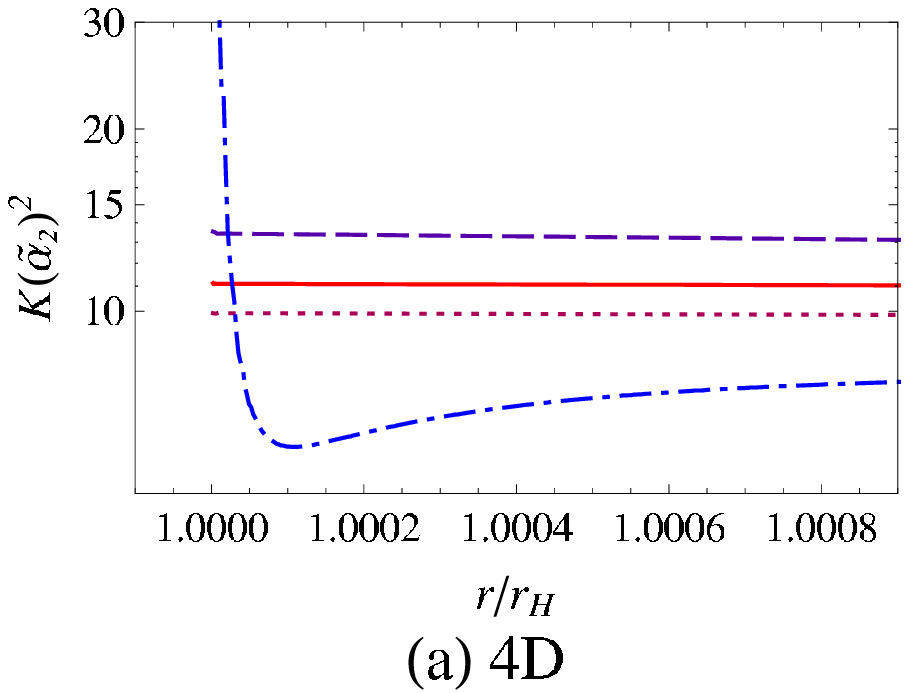}~~~
\includegraphics[width=58mm]{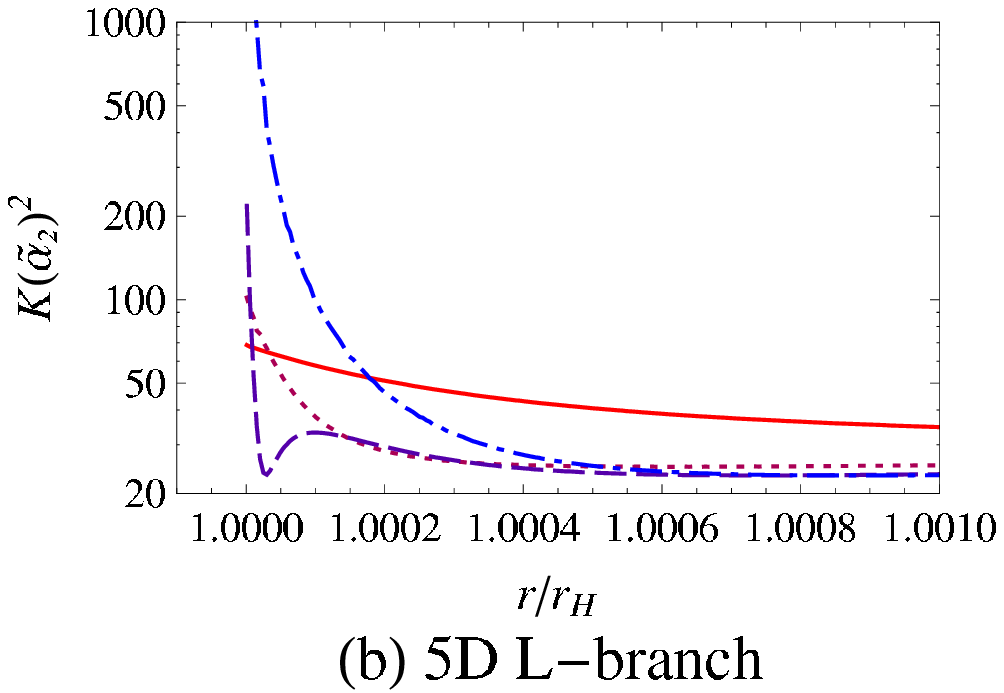}~~~
\includegraphics[width=58mm]{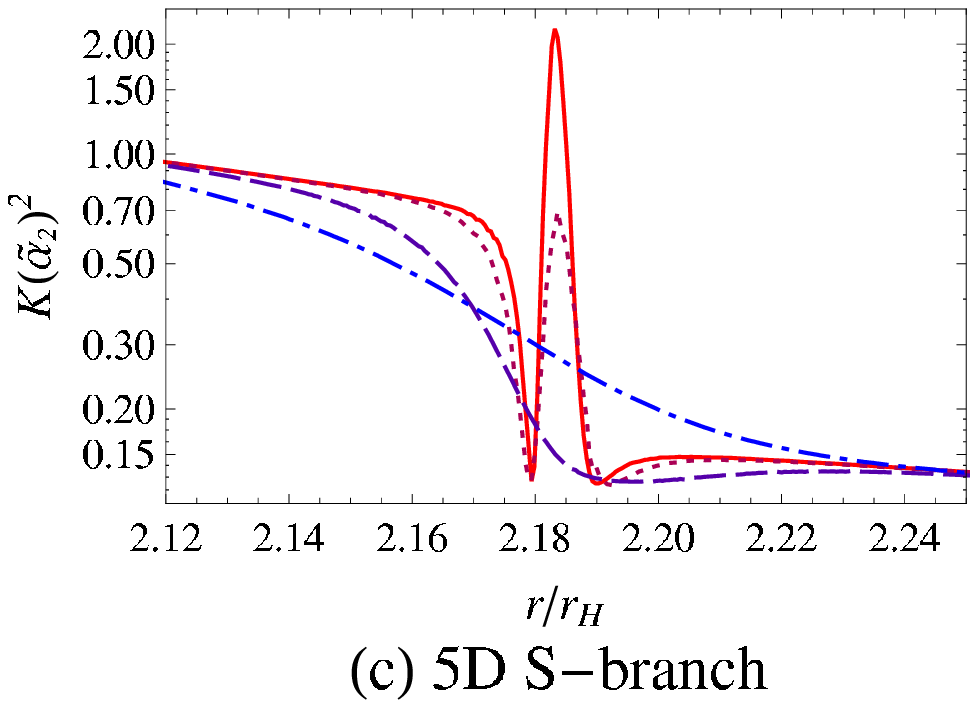}
\caption{The Kretschmann curvature invariants for several horizon radii
in four dimensions and five dimensions.
We choose four values of the horizon radius in four dimensions:
$r_H=$3.66391$\ta_2^{1/2}$ (solid), 3.65865$\ta_2^{1/2}$ (dotted),
3.65751$\ta_2^{1/2}$ (dashed), and 3.65726$\ta_2^{1/2}$ (dot-dashed),
in the L-branch of five dimensions:
$r_H=$2.84950$\ta_2^{1/2}$  (solid), 2.84861$\ta_2^{1/2}$  (dotted),
2.84841$\ta_2^{1/2}$  (dashed), and 2.84837$\ta_2^{1/2}$  (dot-dashed),
and in the S-branch of five dimensions:
$r_H=$0.636663$\ta_2^{1/2}$  (solid), 0.636652$\ta_2^{1/2}$ (dotted),
0.636591$\ta_2^{1/2}$ (dashed), 0.636400$\ta_2^{1/2}$ (dot-dashed).
In four dimensions and the L-branch of five dimensions, the curvature invariant increases rapidly
near the horizon,
and below the minimum radius, we will find a curvature singularity.
On the other hand, in the S-branch of five dimensions,
we find a strange behaviour of the curvature invariant near
$r\approx 2.18 r_H$,
but it does not diverge near the horizon.
}
\label{5D_singularity}
\end{figure}
% \end{widetext}

For the L-branch of the five-dimensional black holes,
we find very similar behaviours to the case of four dimensions.
For the solution with minimum horizon radius, the second derivative
of the dilaton field diverges.
The Kretschmann curvature invariant also diverges at the horizon in the limit of
$M\rightarrow M_{\rm min}^{\rm (L)}$,
as shown in Fig.~\ref{5D_singularity} (b).

%\begin{widetext}

\begin{figure}[ht]
\includegraphics[width=58mm]{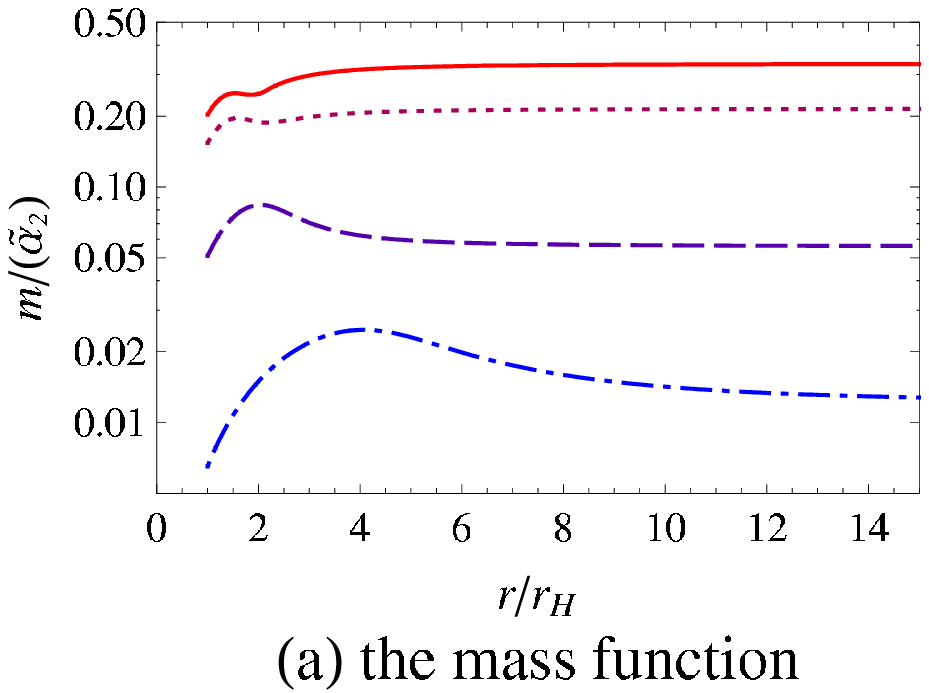}~~~
\includegraphics[width=58mm]{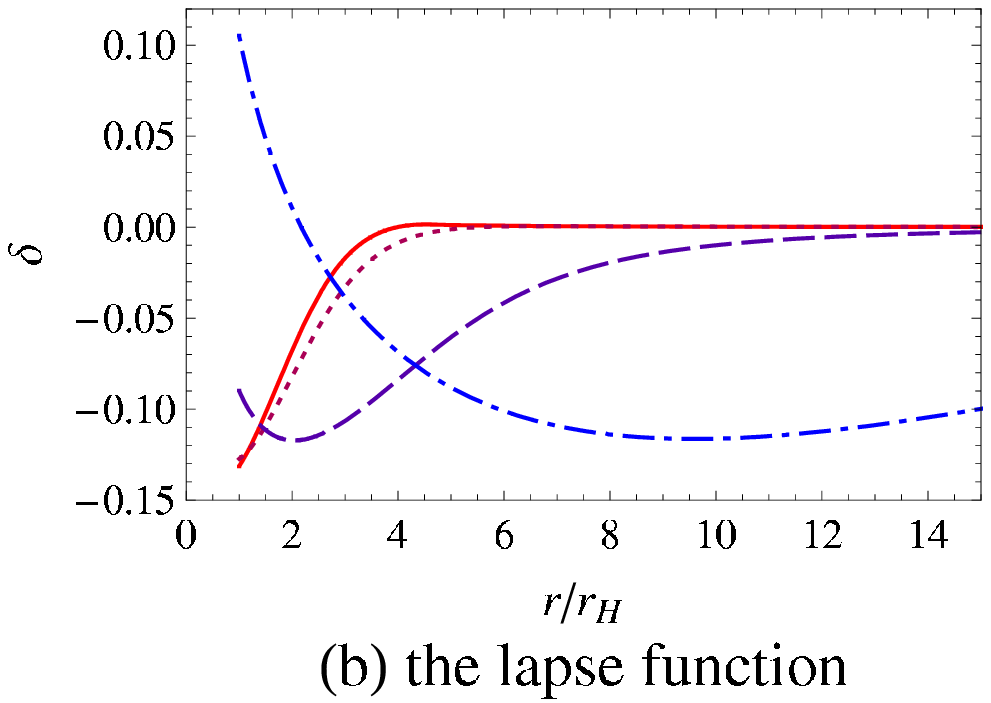}~~~
\includegraphics[width=58mm]{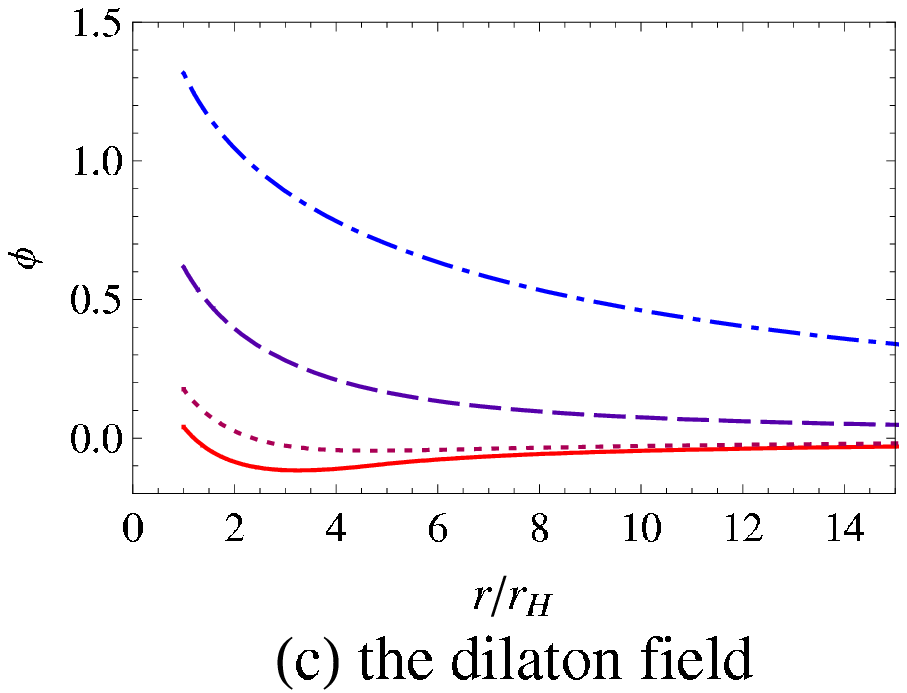}
\caption{
The solution of the S-branch of the five-dimensional black hole.
We choose the following four values for the horizon radius:
 $r_H=$0.636670$\ta_2^{1/2}$  (solid), 0.554630$\ta_2^{1/2}$ (dotted),
 0.319473$\ta_2^{1/2}$ (dashed), 0.114232$\ta_2^{1/2}$ (dot-dashed).
Near the maximum mass of the S-branch, we find some irregular behaviour
just outside of the horizon.
}
\label{5Dsproperties}
\end{figure}
%\end{widetext}

In the S-branch, however, we find somewhat different result.
As we see in Fig. \ref{5Dsproperties}, we find some irregular behaviour
just outside of the horizon ($r\sim 2r_H$) for the black holes with
$r_H$= 0.636670$\ta_2^{1/2}$  (the solid line)
and 0.554630$\ta_2^{1/2}$ (the dotted line), and
the corresponding masses ($M=19.7733\ta_2$ and $12.7352\ta_2$)
are close to the maximum values $M_{\rm max}^{\rm (S)}(=19.7733\ta_2)$
 in the S-branch.
We show the Kretschmann curvature invariant for several radii in
Fig.~\ref{5D_singularity} (c).
Although we find a strange behaviour of the curvature invariant
around $r\approx 2.18 r_H$, the curvature does not seem to diverge
on the horizon.
We suspect that the point with the irregular behaviour
outside the horizon will
become a singularity in the limit of $M
\rightarrow M_{\rm max}^{\rm (S)}$.

%\begin{widetext}
\begin{figure}[ht]
\includegraphics[width=58mm]{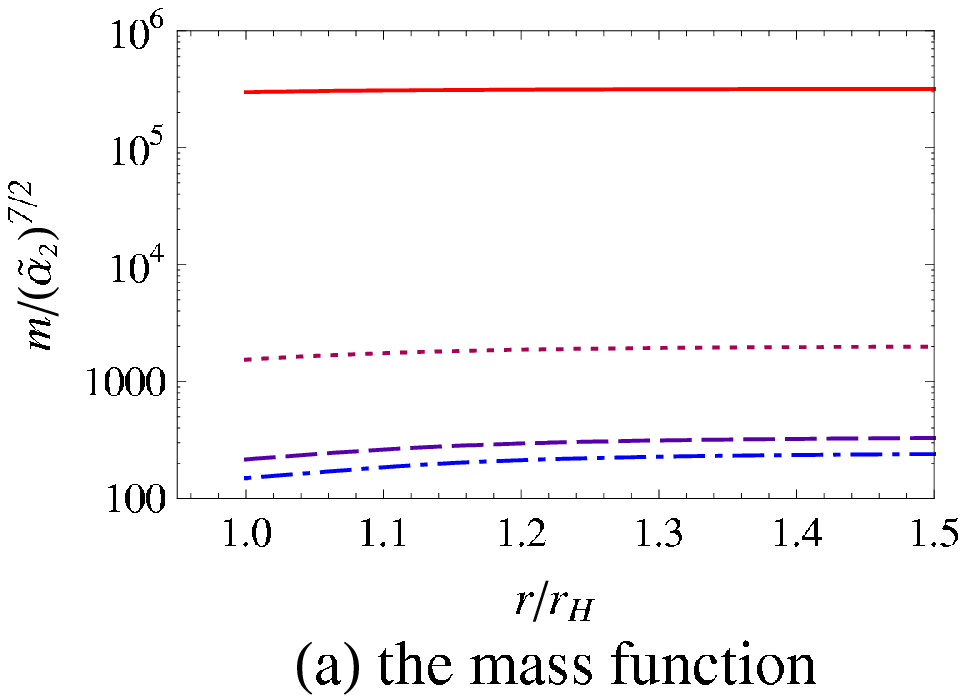}~~~
\includegraphics[width=58mm]{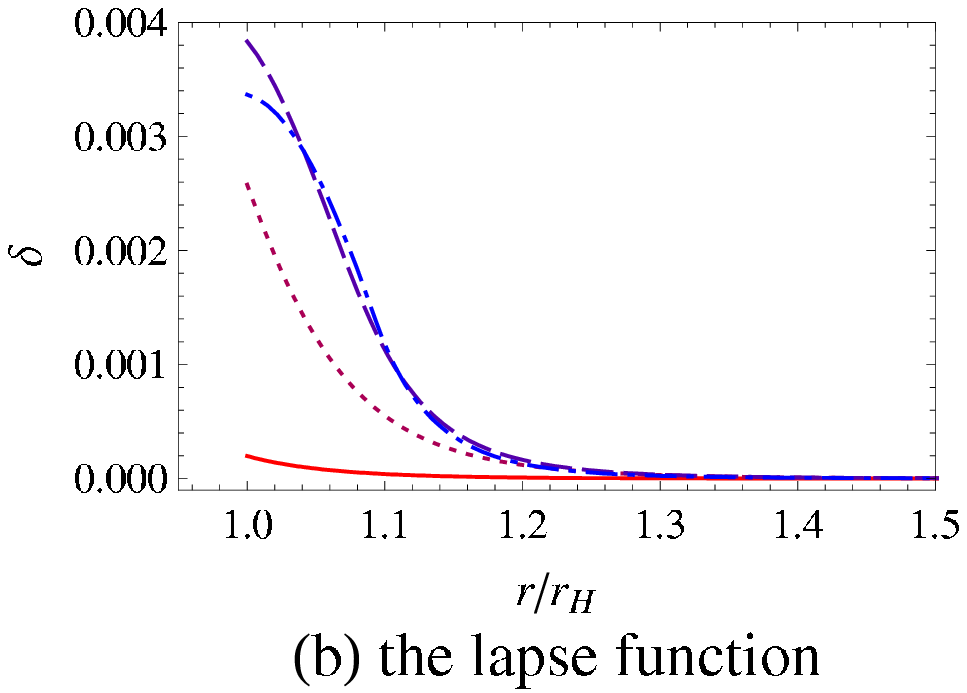}~~~
\includegraphics[width=58mm]{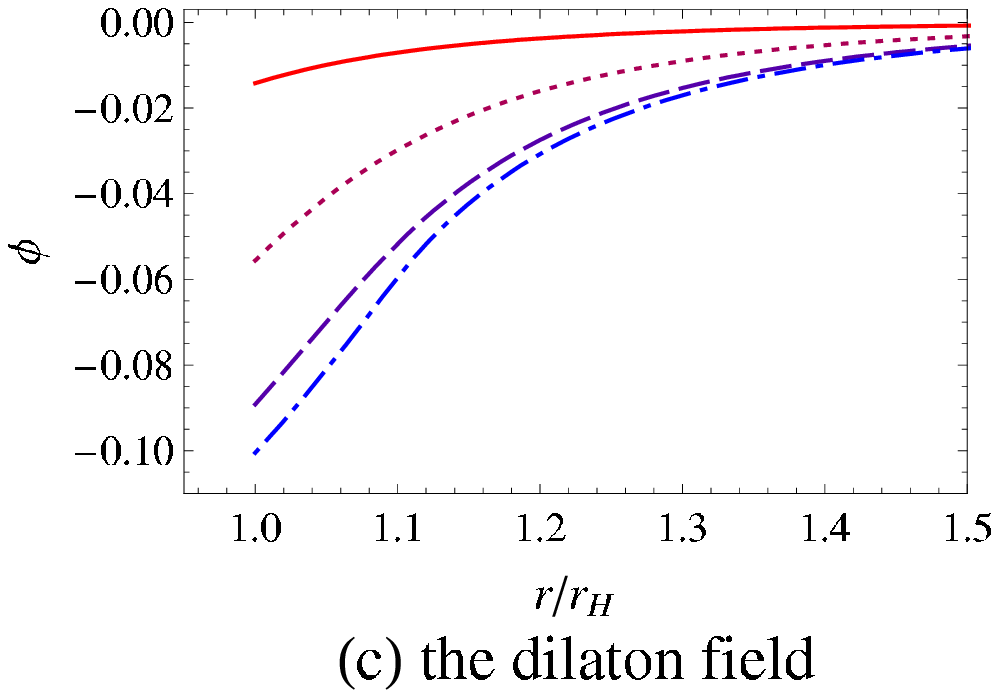}
\caption{The solution of the L-branch of the ten-dimensional black hole.
We choose the following four values for the horizon radius:
 $r_H=$6.683666$\ta_2^{1/2}$  (solid), 3.150271$\ta_2^{1/2}$  (dotted),
2.379547$\ta_2^{1/2}$  (dashed), and 2.258180$\ta_2^{1/2}$  (dot-dashed).
Near the minimum mass of the L-branch, the lapse function and
the dilaton field diverge at the horizon. The singularity appears
on the horizon in the limit of $M\rightarrow M_{\rm min}^{\rm (L)}$.}
\label{10Dlproperties}
\end{figure}
%--figures-------------------------------------------------------------
\begin{figure}[ht]
\includegraphics[width=58mm]{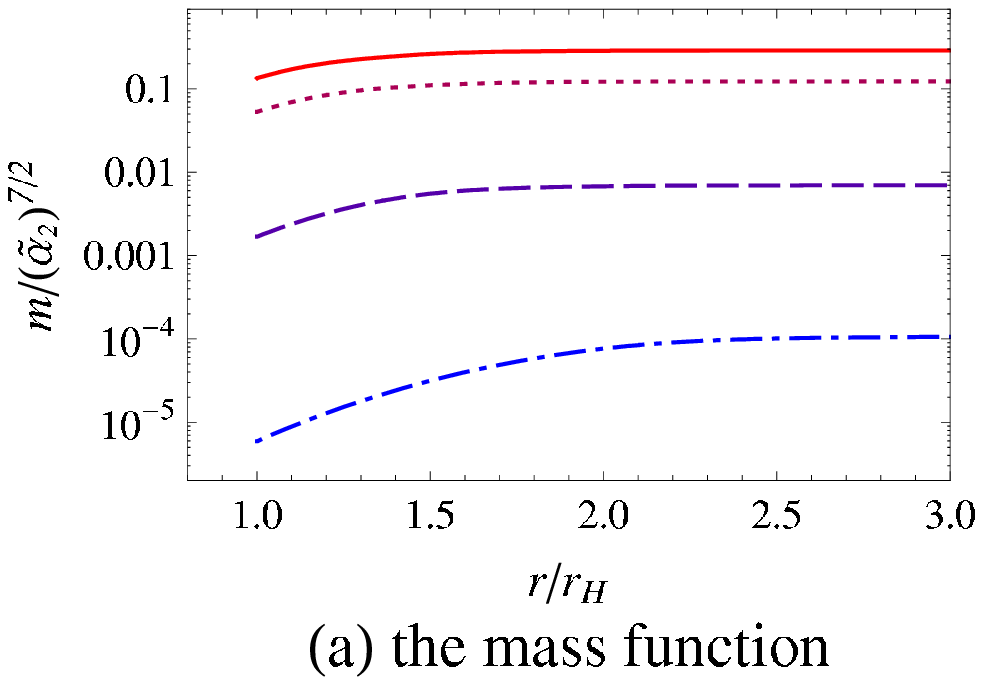}~~~
\includegraphics[width=58mm]{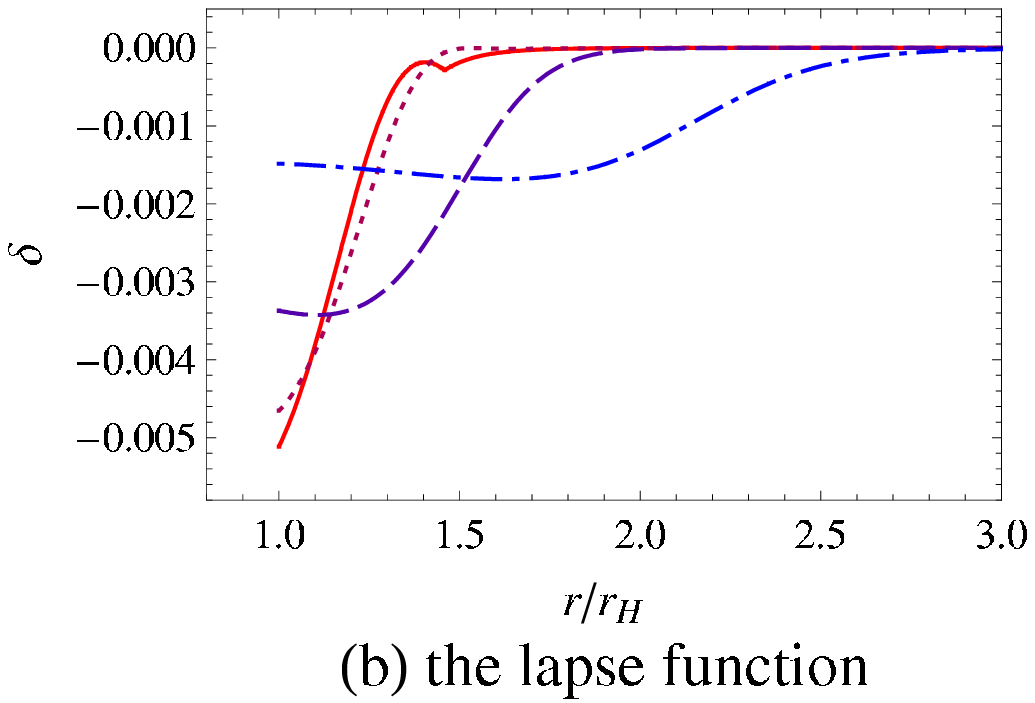}~~~
\includegraphics[width=58mm]{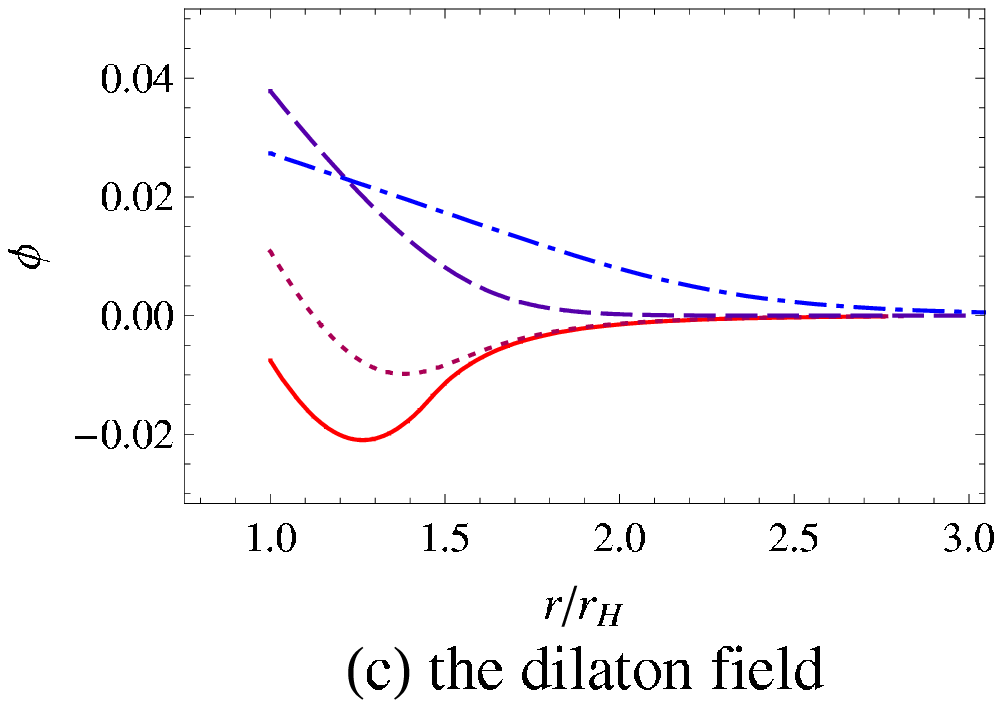}
\caption{The solution of the S-branch of the ten-dimensional black hole.
We choose the following four values for the horizon radius:
$r_H=$0.4135451$\ta_2^{1/2}$ (solid), 0.1639727$\ta_2^{1/2}$ (dotted),
0.0645152$\ta_2^{1/2}$  (dashed), and 0.03157994$\ta_2^{1/2}$ (dot-dashed).
}
\label{10Dsproperties}
\end{figure}

\end{widetext}

\begin{figure}[ht]
\includegraphics[width=60mm]{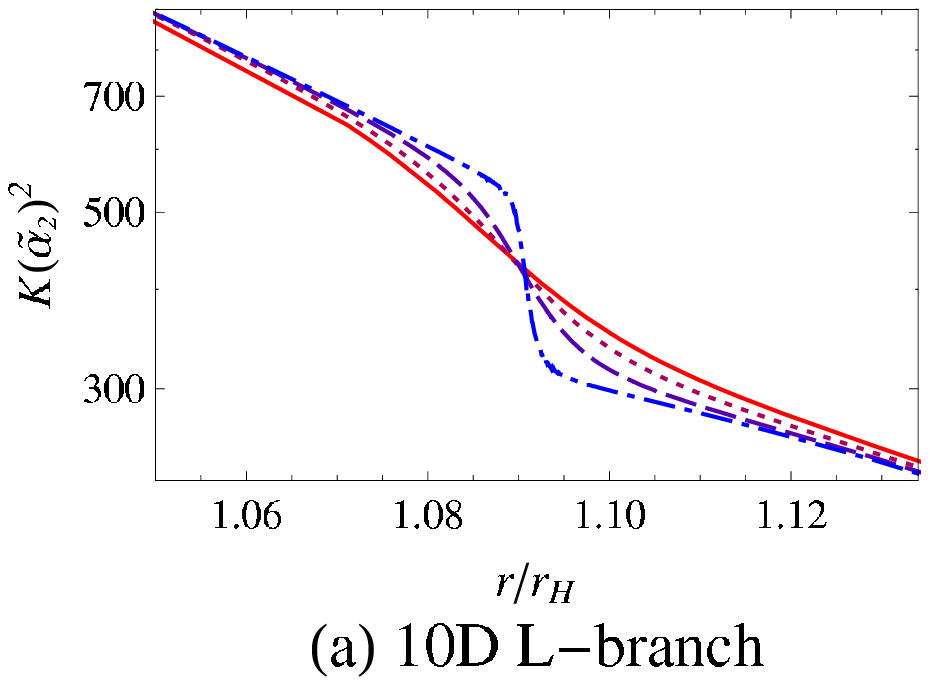}~~~
\\
\vskip .5cm
\includegraphics[width=60mm]{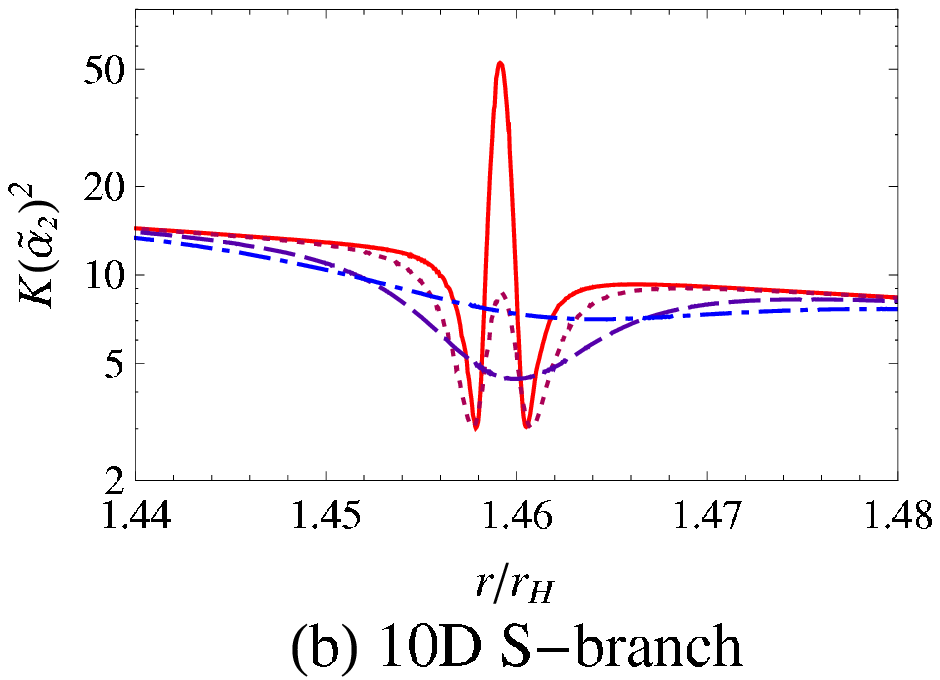}
\caption{The curvature invariants for several masses in ten dimensions.
We choose the following four values for the horizon radius in the L-branch:
$r_H=2.27983\ta_2^{1/2}$  (solid), $2.26830\ta_2^{1/2}$  (dotted),
$2.26116 \ta_2^{1/2}$  (dashed), and $2.25820\ta_2^{1/2}$  (dot-dashed),
In the S- branch:
$r_H=0.828778\ta_2^{1/2}$ (solid), $0.828748\ta_2^{1/2}$ (dotted),
$0.828366\ta_2^{1/2}$  (dashed), and $0.826771\ta_2^{1/2}$ (dot-dashed).
Near the minimum radius in the L- branch, the invariant
does not diverge, but evolves into a ``gravitational
shock wave" at $r\approx 1.09 r_H$ in the limit of minimum radius.
In the S-branch, a strange behaviour appears
near $r\approx 1.46 r_H$ outside the horizon.
It may evolve into a singularity in the limit of
$M\rightarrow M_{\rm max}^{\rm (S)}$.}
\label{10D_singularity}
\end{figure}
% \end{widetext}
Next we depict the mass function, lapse function and dilaton field
in ten dimensionsin Figs. \ref{10Dlproperties} and \ref{10Dsproperties}.
We find the very different behaviour
in the L-branch from the four-dimensional case or from the L-branch in five dimensions.
In the L-branch, the lapse function and the dilaton field
does not diverge near the horizon in the limit of $r_H
\rightarrow r_{\rm min}^{\rm (L)}$.
We show the Kretschmann curvature invariant
in Fig. \ref{10D_singularity} (a).
There appears a discontinuity in the curvature invariant
around $r\approx 1.09r_H$ for $r_H= 2.25818 \ta^{1/2}$.
This point does not evolve into the divergence of the curvature even
in the limit of  $r_H\rightarrow r_{\rm min}^{\rm (L)}$.
We may regard it as a ``gravitational shock wave", where
we have a curvature discontinuity.
It is a new type of singularity.
The reason why we find the minimum radius
in the L-branch is different from the minimum radius found by the regular
horizon condition is that the ``gravitational shock wave" appears
first outside of the horizon
before the singularity appears on the horizon when we take the limit
of $r_H\rightarrow r_{\rm min}^{\rm (L)}$.

In the S-branch, we also find the similar behaviour to
the S-branch in five dimensions. Some irregular behaviour appears
around $r \approx 1.46 r_H$ outside of the horizon
near the maximum mass $M_{\rm max}^{\rm (S)}$.
It may evolve into a singularity in the limit of $M
\rightarrow M_{\rm max}^{\rm (S)}$.

Now we can summarize our results as follows:
For the four dimensions and the L-branch in five dimensions,
there is a minimum radius, below which
the curvature diverges on the horizon.
When $D\geq 6$, the singularity of the gravitational shock-wave
appears in the L-branch below the minimum radius.
On the other hand, for the S-branch of five and higher dimensions,
there exists a maximum mass, beyond which
a black hole does not exist, and a
 singularity may appear outside the horizon.

%%%%%%%%%%%%%%%%%%%%%%%%%%%%%%%%%%%%%%%%%%%%%%%%%%%%%%%%%%%%%%%%%%
%%%%%%%%%%%%%%%%%%%%%%%%%%%%%%%%%%%%%%%%%%%%%%%%%%%%%%%%%%%%%%%%%%
%%%%%%%%%%%%%%%%%%%%%%%%%%%%%%%%%%%%%%%%%%%%%%%%%%%%%%%%%%%%%%%%%%
\section{Truncated Dilatonic Einstein-Gauss-Bonnet model}
\label{sec7}
%%%%%%%%%%%%%%%%%%%%%%%%%%%%%%%%%%%%%%%%%%%%%%%%%%%%%%%%%%%%%%%%%%
%%%%%%%%%%%%%%%%%%%%%%%%%%%%%%%%%%%%%%%%%%%%%%%%%%%%%%%%%%%%%%%%%%
%%%%%%%%%%%%%%%%%%%%%%%%%%%%%%%%%%%%%%%%%%%%%%%%%%%%%%%%%%%%%%%%%%
In this section, we discuss the truncated Einstein-Gauss-Bonnet
model, whose action is given by Eq.~(\ref{Taction}).
The properties of black hole solutions in the TDEGB model with $\c=\frac{1}{2}$ are
studied in~\cite{ohta_torii1}.
This value of the coupling constant is obtained in ten dimensions.
If we start from the effective action in the string frame in $D$ dimensions,
we have the different value, which is  $\c=\sqrt{2/(D-2)}$,
in the Einstein frame.
Here we choose this value of the coupling constant for the TDEGB model.
The difference of two models originates from the compactification.
We find some qualitative differences in thermodynamical properties
in the TDEGB models with these two coupling constants
(see the discussion in Sec.~\ref{sec9}).
In what follows in the text, we show the black hole solutions
and their properties in the TDEGB model with $\c=\sqrt{2/(D-2)}$.

%%%%%%%%%%%%%%%%%%%%%%%%%%%%%%%%%%%%%%%%%%%%%%%%%%%%%%%%%%%%%%%%%%
%%%%%%%%%%%%%%%%%%%%%%%%%%%%%%%%%%%%%%%%%%%%%%%%%%%%%%%%%%%%%%%%%%
\subsection{Basic equations}
%%%%%%%%%%%%%%%%%%%%%%%%%%%%%%%%%%%%%%%%%%%%%%%%%%%%%%%%%%%%%%%%%%
%%%%%%%%%%%%%%%%%%%%%%%%%%%%%%%%%%%%%%%%%%%%%%%%%%%%%%%%%%%%%%%%%%
The field equations are given by
\begin{widetext}
\begin{align}
fr^2 F_{{\rm T}(\phi)}&:=
\tfrac{1}{2}\lh f'r +h +2(D-2)f \rh \phi' r + f \phi'' r^2
 \nn
&-\c B \bigl[
(D-4)_{5} (1-f)^2 - 2 (D-4)(1-f)(f'r +h) +4 X f (1-f) r^2 +2 h f' r
 \bigr]=0 , \\
fr^2 F_{{\rm T}(\nu)}&:=
(D-2)_3 (1-f) -(D-2)f'r -\tfrac{1}{2} f \phi'^2 r^2
\nn
&+ B \bigl[
(D-4)_{5} (1-f)^2 -2 (D-4) (1-f) (f' - 2 \c f \phi' )r
\nn
&\hspace{8mm}
 +4 \c f (1-f) ( \phi'' - \c \phi'^2 ) r^2 +2 \c (1-3f) \phi' f' r^2
 \bigr]=0 , \\
fr^2 F_{{\rm T}(\la)}&:=
(D-2)_3 (1-f) -(D-2)h +\tfrac{1}{2} f \phi'^2 r^2 \nn
&+ B \bigl[
(D-4)_{5} (1-f)^2 -2 (D-4)(1-f) ( h - 2 \c f \phi' r ) +2 \c h (1-3f)
\phi' r \bigr]=0 , \\
fr^2 F_{{\rm T}(\mu)}&:=
(D-2)_4 (1-f) -(D-2)_3 (f'r +h )
 +2 (D-2) X fr^2  -\tfrac{1}{2}(D-2) f \phi'^2 r^2 \nn
&+ B \bigl[
(D-4)_{6} (1-f)^2 -2 (D-4)_5 (1-f) (f'r +h) +4 \c (D-4)_5 f (1-f) \phi' r \nn
& \hspace{8mm} +4 (D-4) (1-f)X f r^2 +4 \c (D-4)  f (1-f)
( \phi'' - \c \phi'^2 ) r^2
 +2(D-4) h f'r \nn
& \hspace{8mm}
 +2 \c (D-4)  (1-3f) (f'r +h) \phi' r
-4 \c f h ( \phi'' - \c \phi'^2 ) r^2 -4 \c h \phi' f' r^2 + 8 \c \phi'
X f^2 r^3
 \bigr] =0
\,.
\end{align}
where
% \begin{align}
% X (r):=\frac{1}{4f^2r^2} \left[ h(f'r-h) -2f ( h'r - h ) \right]
% \,,
% &
% B(r) := r^{-2} \ta_2 e^{-\c \phi}
% \,,
% &
% h(r) := r(f'-2f\d')
% \,.
% \end{align}
\begin{eqnarray}
X (r)&:=&\frac{1}{4f^2r^2} \left[ h(f'r-h) -2f ( h'r - h ) \right]
\,,
\\
B(r) &:= & r^{-2} \ta_2 e^{-\c \phi}
\,,
\\
h(r) &:= &r(f'-2f\d')
\,.
\end{eqnarray}
\end{widetext}

The Bianchi identity gives one relation between these
four functionals:
\begin{eqnarray}
&&
f^{-1/2}\left(f^{1/2} F_{{\rm T}(\la)}\right)'
\nonumber \\
&&~~
={1\over r}F_{{\rm T}(\mu)}+
\left({f'\over 2f}-\delta'\right)
F_{{\rm T}(\nu)}
+\phi'F_{{\rm T}(\phi)}
\,.
% \nn
~~
\end{eqnarray}
Hence if one solve three of them, the remaining one equation
is automatically satisfied.
%\end{widetext}

%%%%%%%%%%%%%%%%%%%%%%%%%%%%%%%%%%%%%%%%%%%%%%%%%%%%%%%%%%%%%%%%%%
%%%%%%%%%%%%%%%%%%%%%%%%%%%%%%%%%%%%%%%%%%%%%%%%%%%%%%%%%%%%%%%%%%
\subsection{Boundary conditions}
\label{boundary_TDEGB}
%%%%%%%%%%%%%%%%%%%%%%%%%%%%%%%%%%%%%%%%%%%%%%%%%%%%%%%%%%%%%%%%%%
%%%%%%%%%%%%%%%%%%%%%%%%%%%%%%%%%%%%%%%%%%%%%%%%%%%%%%%%%%%%%%%%%%
As we discussed in Sec.~\ref{boundary_DEGB}, we
need the boundary conditions
both at the event horizon ($r=r_H$) and at the infinity
($r=\infty$).
The asymptotical ``flatness" implies
\begin{eqnarray}
f &\rightarrow& 1-\left[{2\kappa_D^2
 \over (D-2)A_{D-2}}\right]{M\over r^{D-3}}
\,,
\nonumber
\\
\delta
&\rightarrow& {\cal O}\left({1\over r^{2(D-3)}}\right)
\,,~~~~~~~~~~~~~~~~~~~~
\nonumber
\\
\phi &\rightarrow& {\cal O}\left({1\over r^{D-3}}\right)
\,,~~~~~~~~~~~~
\end{eqnarray}
as $r\rightarrow \infty$.
$M$ is a gravitational mass in the Einstein frame.
Since the weak equivalence principle is satisfied in the Einstein frame,
the lapse must drop faster than the gravitational potential $(f-1)$.

\vskip .5cm
\begin{table}[ht]
\caption{The allowed values for the regular horizon radius are shown.
The equality gives a double root of $\phi_H'$.
There is a minimum radius in four dimensions. In five dimensions and six dimensions, there are gaps in which there is
no regular horizon.
For dimensions higher than six, there is a regular horizon for any
horizon radius.
$\bar \rho_H:= r_He^{\c\phi_H/2}/\ta_2^{1/2}$.}
\begin{center}
\begin{tabular}{|c|l|}
\hline
$D$& ~~Condition for
regular horizon
\\
\hline
\hline
4&~~$\bar \rho_H \geq {1.86121}  $~~
\\
\hline
5&~~$\bar \rho_H \geq {1.45828} \,~~
{\rm or}~~~\bar \rho_H \leq  {0.962882} $~~
\\
\hline
~$6\leq D\leq 10$~& ~~any values
\\
\hline
\end{tabular}
\label{table_3}
\end{center}
\end{table}
\noindent
At the event horizon ($r_H$),
the metric function $f$ vanishes,
i.e.
$f(r_H)= 0
$\,.
 The variables and their derivatives must be finite at $r_H$.

\begin{widetext}
The regularity conditions of the event horizon are now
\begin{eqnarray}
&&
\bar \rho_H^2\xi_H\eta_H-\c 
\left[(D-4)_5 -4(D-4)\xi_H+4\zeta_H+2\xi_H^2\right]=0,
\label{BCE1}
\\
&&
\bar \rho_H^2\left[(D-2)_3-(D-2)\xi_H\right]
+
\left[(D-4)_5 -2(D-4)\xi_H+2\c \xi_H\eta_H\right]=0,
\label{BCE2}
\\
&&
\bar \rho_H^2\left[(D-2)_4-2(D-2)_3\xi_H+2(D-2)\zeta_H \right]
\nn
&&
~~+
[(D-4)_6 -4(D-4)_5\xi_H+4(D-4)\zeta_H
+4\c (D-4)\xi_H\eta_H+2(D-4)\xi_H^2-4\c \xi_H^2\eta_H
]=0,~~~~
\label{BCE3}
\end{eqnarray}
where
\begin{eqnarray}
\bar \rho_H:=
r_He^{\c\phi_H/2}/\ta_2^{1/2}
~,~~
\xi_H:= r_H f_H'~,~~
\eta_H:=  r_H \phi_H'~,~~{\rm and}~~
\zeta_H:=  r_H^2 (X f)_H
\,.
\end{eqnarray}
Eliminating $\xi_H$ and $\zeta_H$ in Eqs.
(\ref{BCE1}) $\sim$ (\ref{BCE3})
[we assume that $\xi_H\neq 0$ and $\zeta_H \neq 0$], we find the
 quadratic equation for $\eta_H$:
\begin{eqnarray}
a\eta_H^2+b\eta_H+c=0
\label{eq_phip2}
\,,
\end{eqnarray}
where
\begin{eqnarray}
a&=&2\c
\Bigl[ ((D-2)_3 \bar \rho_H^2 + (D-4)_5 )( (D-2) \bar \rho_H^2 +2(D-4) )
+2\c^2 ((D-2) (D-4) (3D-11)\bar \rho_H^2 +4(D-3) _5 ) \Bigr],
\nn
b&=&
- \left[
 \bigl( (D-2)_3 \bar \rho_H^2 +(D-4)_5 \bigr)
 \bigl( (D-2) \bar \rho_H^2 +2 (D-4) \bigr)^2
\right.
\nn
&&
\left.
 -4 \c^2 (D-1) (D-4) \left( (D-2)^2 \bar \rho_H^4
+2 (D-2) \bar \rho_H^2 -2 (D-4)_5 \right)
\right],
\nn
c&=&\c (D-1)_2 \bar \rho_H^2 \left[ (D-2)^3 \bar \rho_H^4
 -4 (D-2) (D-4) \bar \rho_H^2 -2 (D+1) (D-4)^2 \bigr)
\right]
\,,
\end{eqnarray}
\end{widetext}
For $D=4\sim 10$,
assuming $\ta_2>0$  and imposing the discriminant is non-negative
(${\cal D}_D:=b^2-4ac \geq 0$), we find the
allowed values for the regular event horizon, which are summarized in Table
\ref{table_3}.
There is a minimum horizon radius
$ r_H = 1.86121 ~\ta_2^{1/2}e^{-\c\phi_H/2}$
in four-dimensional spacetime,
while in five-dimensional spacetime, there is a small gap in the parameter space of
horizon radius
($0.962882 ~\ta_2^{1/2}e^{-\c\phi_H/2}
< r_H< 1.45828 ~\ta_2^{1/2}e^{-\c\phi_H/2} $)
where there is no regular horizon. For higher dimensional spacetime
than five dimensions, there is a regular horizon for any
horizon radius.

%%%%%%%%%%%%%%%%%%%%%%%%%%%%%%%%%%%%%%%%%%%%%%%%%%%%%%%%%%%%%%%%%%
%%%%%%%%%%%%%%%%%%%%%%%%%%%%%%%%%%%%%%%%%%%%%%%%%%%%%%%%%%%%%%%%%%
\section{Comparison with TDEGB and EGB}
\label{sec8}
%%%%%%%%%%%%%%%%%%%%%%%%%%%%%%%%%%%%%%%%%%%%%%%%%%%%%%%%%%%%%%%%%%
%%%%%%%%%%%%%%%%%%%%%%%%%%%%%%%%%%%%%%%%%%%%%%%%%%%%%%%%%%%%%%%%%%
Now we  compare the properties of the black hole solutions in the DEGB
and TDEGB models.
We also show the results in the EGB model as a reference.
% \end{widetext}
%%%%%%%%%%%%%%%%%%%%%%%%%%%%%%%%%%%%%%%%%%%%%%%%%%%%%%%%%%%%%%%%%%
%%%%%%%%%%%%%%%%%%%%%%%%%%%%%%%%%%%%%%%%%%%%%%%%%%%%%%%%%%%%%%%%%%
\subsection{Mass-area relation}
%%%%%%%%%%%%%%%%%%%%%%%%%%%%%%%%%%%%%%%%%%%%%%%%%%%%%%%%%%%%%%%%%%
%%%%%%%%%%%%%%%%%%%%%%%%%%%%%%%%%%%%%%%%%%%%%%%%%%%%%%%%%%%%%%%%%%
First we show the mass-area relations of black holes in Fig.~\ref{MA}.
The solid (red) line, dashed (green) line, and dot-dashed (black) line
correspond to the DEGB, TDEGB, and EGB models, respectively.

In four dimensions, there exists the minimum finite radius,
$r_{H({\rm min})}=3.65726 \ta_2^{1/2}$ and $3.138439 \ta_2^{1/2}$,
 and minimum mass,
 $M_{\rm min}^{\rm (DEGB)}=69.3511 \ta_2^{1/2}$ and
$M_{\rm min}^{\rm (TDEGB)} = 42.7128 \ta_2^{1/2}$,
both for the DEGB and TDEGB models,
respectively.
There is no qualitative difference. A turning point appears
at the minimum mass, near where two black holes exist
in a small mass range ($M_{\rm min}^{\rm (DEGB)}\leq M < 72.3945 \ta_2^{1/2}$
for DEGB and $M_{\rm min}^{\rm (TDEGB)}\leq M <42.7152\ta_2^{1/2}$
for TDEGB).
In the EGB model, it is just a Schwarzschild black hole because the Gauss-Bonnet term
is a total divergence. It is completely different from the other two.

In five dimensions, the result changes drastically (see Fig.~\ref{MA}(b)).
For the EGB model, the mass-area relation is very simple
(the dot-dashed line).
The area increases monotonically with respect to the mass,
and there exists a minimum finite mass
($M_{\rm min}^{\rm (EGB)}=2\pi^2 \ta_2$).
In the TDEGB model, the mass-area relation is also monotonic,
but it splits up into two branches (the S- and L-branches)
shown by the dashed (green) line just as in the DEGB model (the solid [red] line).
The ranges are $M_{\rm min}^{\rm (S)}<M<M_{\rm max}^{\rm (S)}$ and
$M>M_{\rm min}^{\rm (L)}$ for the S- and L-branches, respectively,
where
$M_{\rm min}^{\rm (S)}:=10.8051 \ta_2$,
$M_{\rm max}^{\rm (S)}:=27.4615 \ta_2$, and
$M_{\rm min}^{\rm (L)}:=142.382\ta_2$.
So there exists a finite minimum mass ($M_{\rm min}^{\rm (S)}$),
and the area in the L-branch increase monotonically without any turning point
as well as that in the S-branch.
However, in the case of the DEGB model, as we discussed
in Sec.~\ref{sec6}, the black holes exist from zero mass, and there appears
a mass gap ($M_{\rm max}^{\rm (S)}=
19.7733\ta_2<M<M_{\rm min}^{\rm (L)}=395.862\ta_2$)
where no black hole exists.
In the L-branch of the DEGB model,
we find two black hole solutions
with the different horizon radii but with the same mass
in the range of $M_{\rm min}^{\rm (L)}<M<395.880 \ta_2 $.
The smaller-radius black hole may be unstable
because the entropy is also smaller.

In ten dimensions, the mass-area relations are almost the same for the
DEGB, TDEGB, and EGB models.
However there is a qualitative difference between the DEGB model and
the TDEGB (or EGB) model.
In the TDEGB (or EGB) model, the area increases monotonically
with respect to the mass
from zero to infinity. The mass vanishes at zero area.
However, in the DEGB model,
there exists a gap in the range of mass ($M_{\rm max}^{\rm (S)}
= 68.6614 \ta_2^{7/2}
\leq M<M_{\rm min}^{\rm (L)}=58647.5 \ta_2^{7/2}$),
and a turning point appears near the minimum mass $M_{\rm min}^{\rm (L)}$.
There exist two different solution with the same mass,
in the range of mass ($M_{\rm min}^{\rm (L)} < M < M_{\rm min}^{\rm (L)}
+ 1.56019 \times 10^{-5} \ta_2^{7/2}$).
The behaviour is the same as that in five dimensions.
\begin{widetext}

We also find the same behaviours in six to nine dimensions.

%--figures-------------------------------------------------------------
\begin{figure}[ht]
\includegraphics[width=58mm]{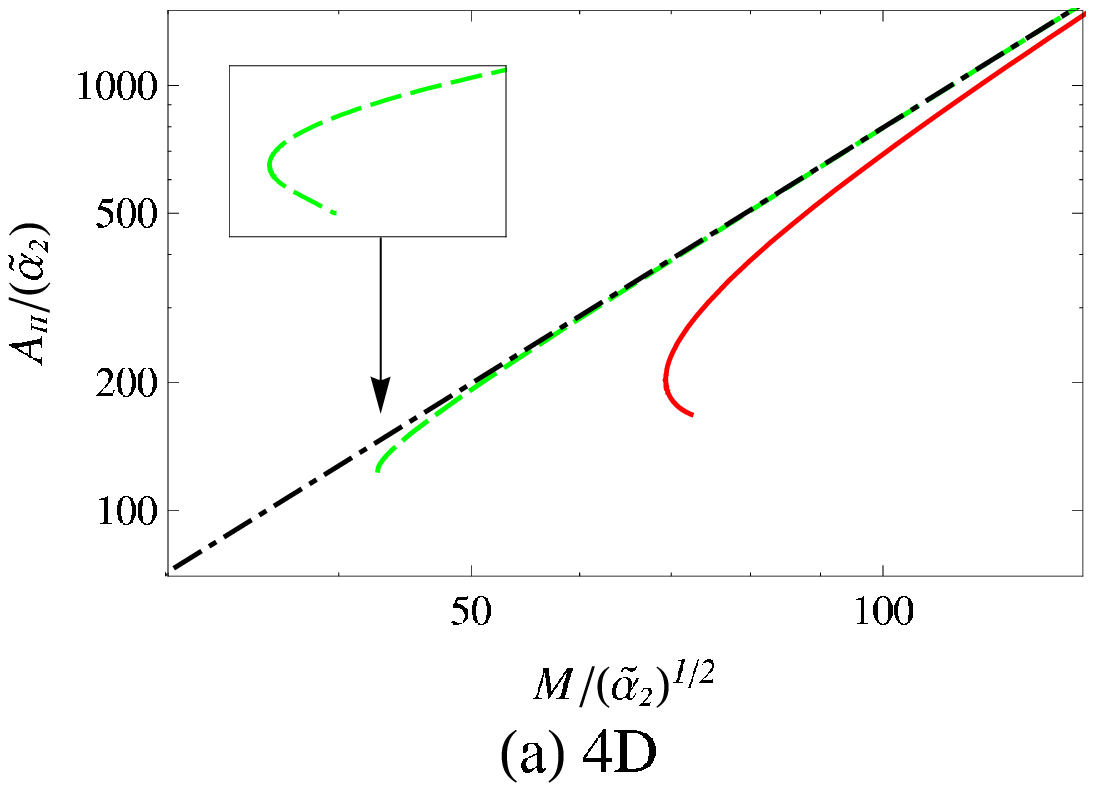}~~~
\includegraphics[width=58mm]{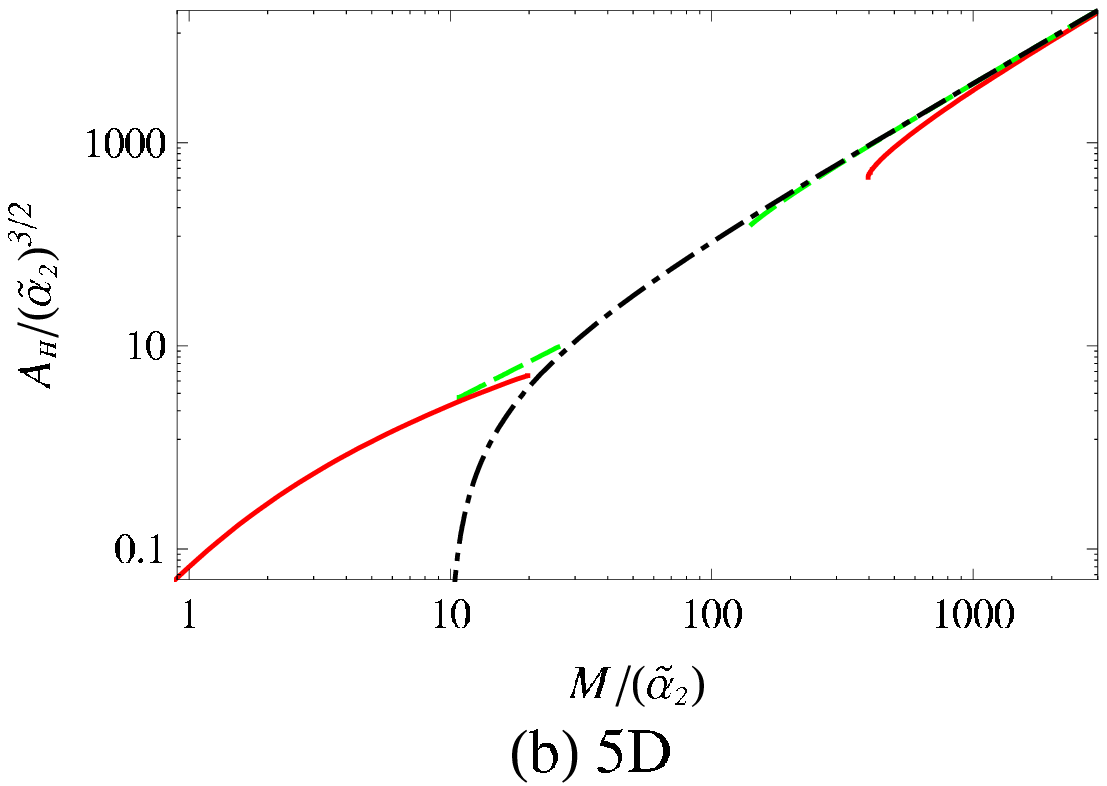}~~~
\includegraphics[width=58mm]{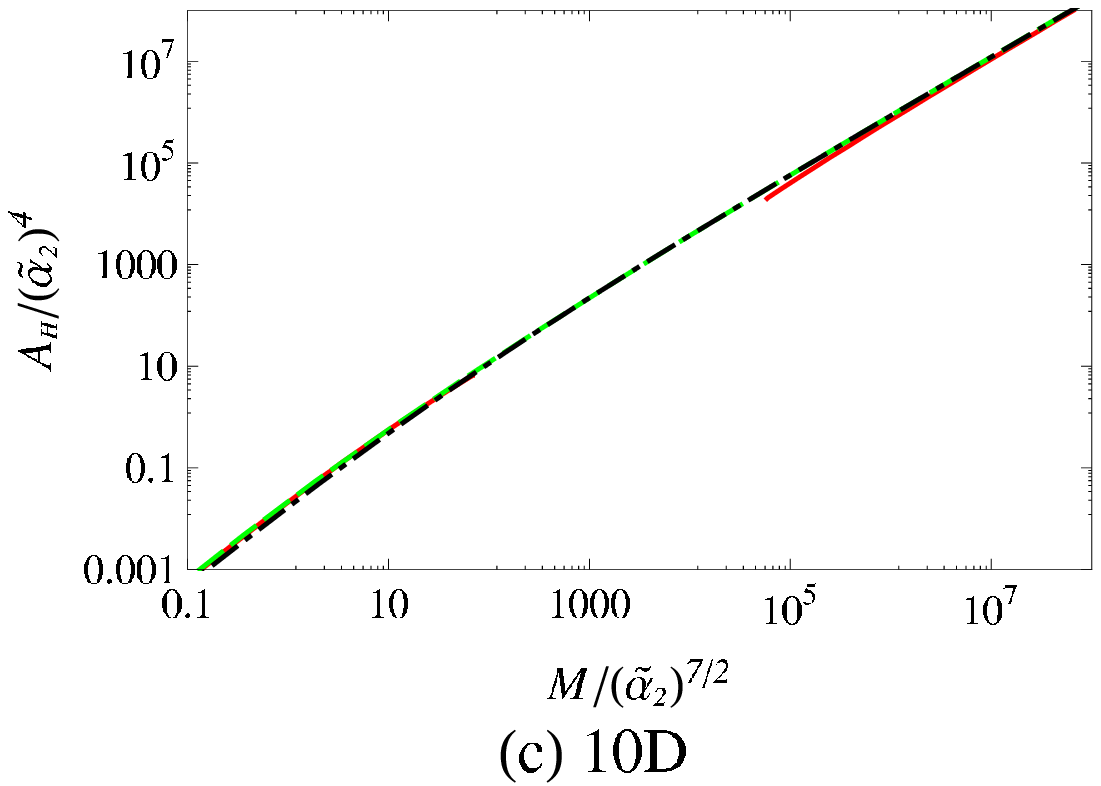}
\caption{The mass-area relations of black hole solutions
in four dimensions, five dimensions, and ten dimensions.
The dashed (green), dot-dashed (black), and solid (red) lines
are the cases of the TDEGB, EGB, and DEGB models, respectively.
}
\label{MA}
\end{figure}
%%--figures-------------------------------------------------------------

\end{widetext}
%%%%%%%%%%%%%%%%%%%%%%%%%%%%%%%%%%%%%%%%%%%%%%%%%%%%%%%%%%%%%%%%%%
%%%%%%%%%%%%%%%%%%%%%%%%%%%%%%%%%%%%%%%%%%%%%%%%%%%%%%%%%%%%%%%%%%
\subsection{Thermodynamics}
%%%%%%%%%%%%%%%%%%%%%%%%%%%%%%%%%%%%%%%%%%%%%%%%%%%%%%%%%%%%%%%%%%
%%%%%%%%%%%%%%%%%%%%%%%%%%%%%%%%%%%%%%%%%%%%%%%%%%%%%%%%%%%%%%%%%%
For the TDEGB model (\ref{Taction}),
the entropy is given by
\begin{align}
S_{\rm T} = \frac{A_H}{4} \lh 1 +\frac{2\ta_2}{r_H^2} e^{-\c \phi_H} \rh,
\end{align}
The corrections from the
Bekenstein-Hawking entropy is
\begin{align}
S_{\rm T}-S_{\rm BH}
=\frac{2\ta_2}{r_H^2}e^{-\c \phi_H}
\times S_{\rm BH}>0
\label{entropy_TDEGB}
\,.
\end{align}
This means that the entropy is always larger than the
Bekenstein-Hawking's one in the TDEGB model as well.
The difference between (\ref{entropy_string}) and (\ref{entropy_TDEGB})
comes from the truncated term ${\cal F}$ (note that the values of $\phi_H$
in the DEGB theory and the truncated one are different).

Here, we show the black hole entropy and temperature in Figs.~\ref{entropy}
and \ref{temperature}, respectively. In these figures,
the solid (red) line, dashed (green) line, and dot-dashed (black) line
describe the results for the  DEGB, TDEGB, and EGB  models, respectively.

The entropy behaves quite similarly to the area for all models,
although the values are slightly different as shown
in Fig.\ref{fig_entropy_DEGB} in the DEGB model.
There is no qualitative difference between $A_H$ and $S$,
except for $D=4$ in the TDEGB model for which
we find a cusp near $M\approx M_{\rm min}^{\rm (TDEGB)}$
instead of a turn-around smooth curve~\cite{TM}.
The cusp is related to a stability change
understood by  a catastrophe theory~\cite{MTTM}.

As for the temperature,
the behaviours are quite different in each model
depending on the dimensions.
In four dimensions, just as the area or the entropy,
there appears a turning point,
at which stability changes
as we expected~\cite{Katz}.
The same behaviour is found in the TDEGB model.
In the EGB model, it is just a Schwarzschild black hole,
i.e., $T_H\propto 1/M$.

In five dimensions, however, there is a maximum temperature
$T_{\rm max}=0.251938 \ta_2^{-{1\over 2}}$
at $M= 15.0506 \ta_2$
in the DEGB model, just as in the EGB model
($T_{\rm max}= {\sqrt{6} \over 8 \pi}\ta_2^{-1/2}$
at $M={4\pi^2}\ta_2$), although we have a mass gap.
The temperature in the TDEGB model decreases
monotonically as the mass increases.

\begin{widetext}

\begin{figure}[ht]
\includegraphics[width=58mm]{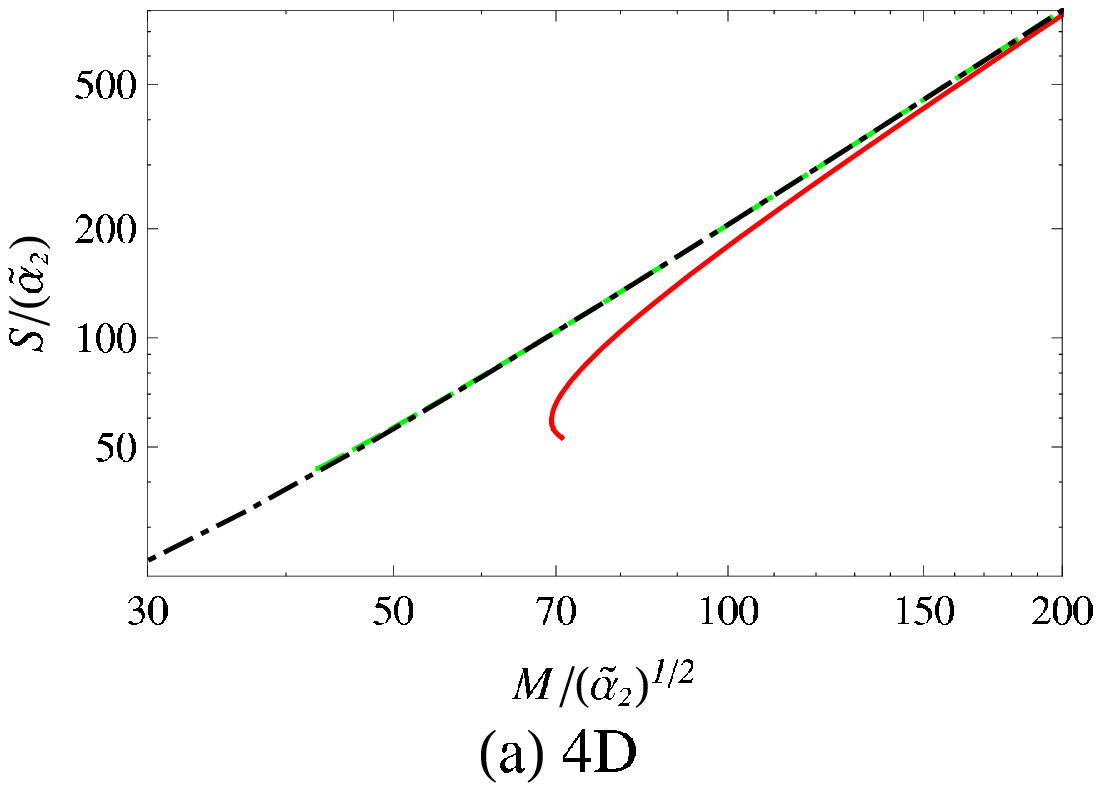}~~~
\includegraphics[width=58mm]{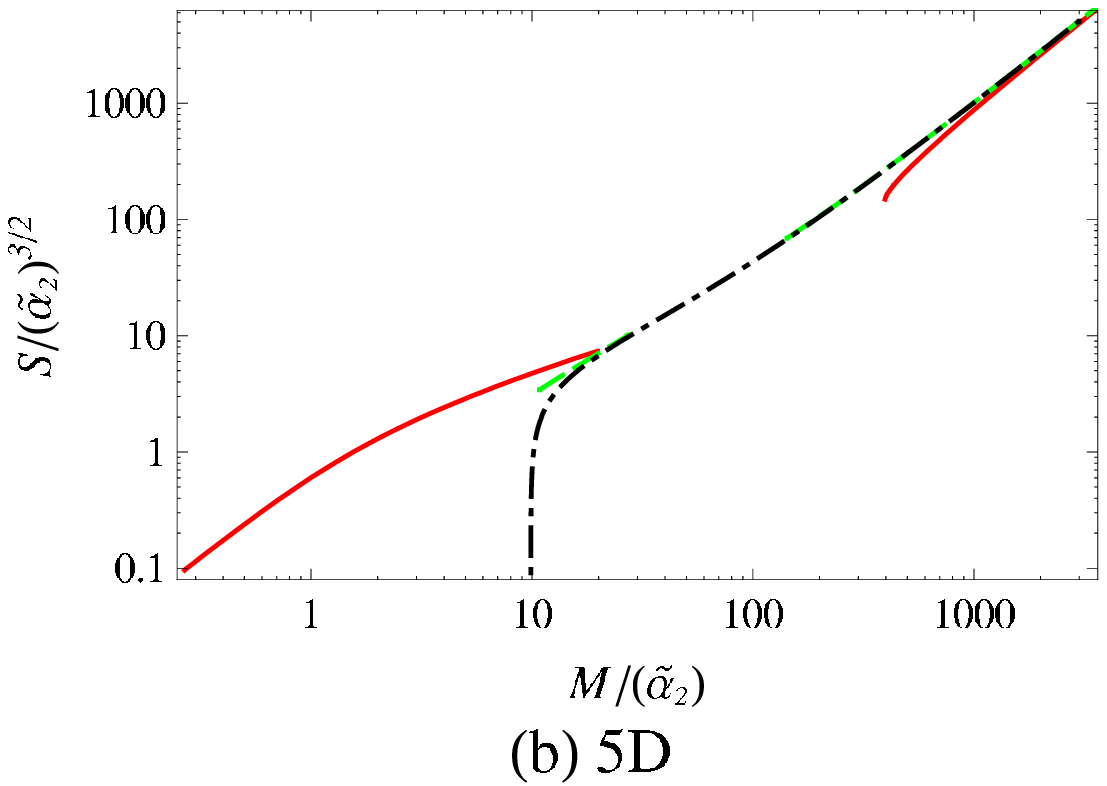}~~~
\includegraphics[width=58mm]{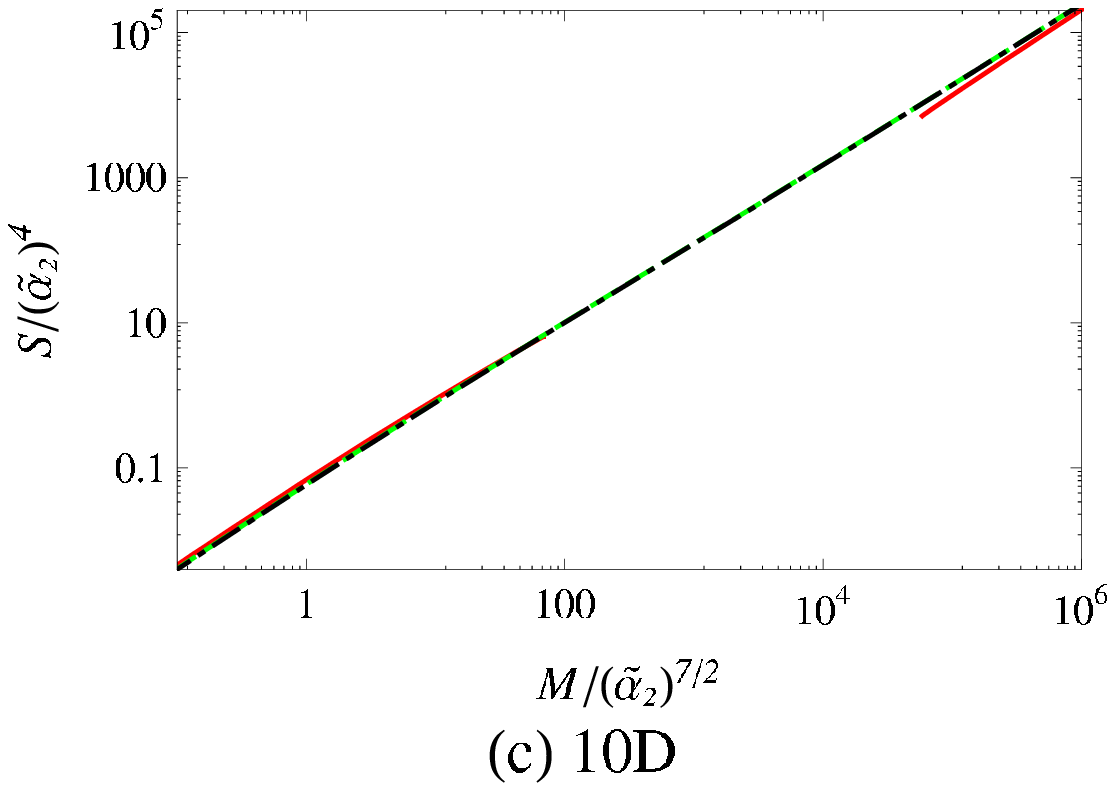}
\caption{The entropies  of black holes with respect to the gravitational mass.
The  solid (red),  dashed (green), and  dot-dashed (black) lines
are for the DEGB, TEGBD, and EGB models.
}
\label{entropy}
\end{figure}
\begin{figure}[ht]
\includegraphics[width=58mm]{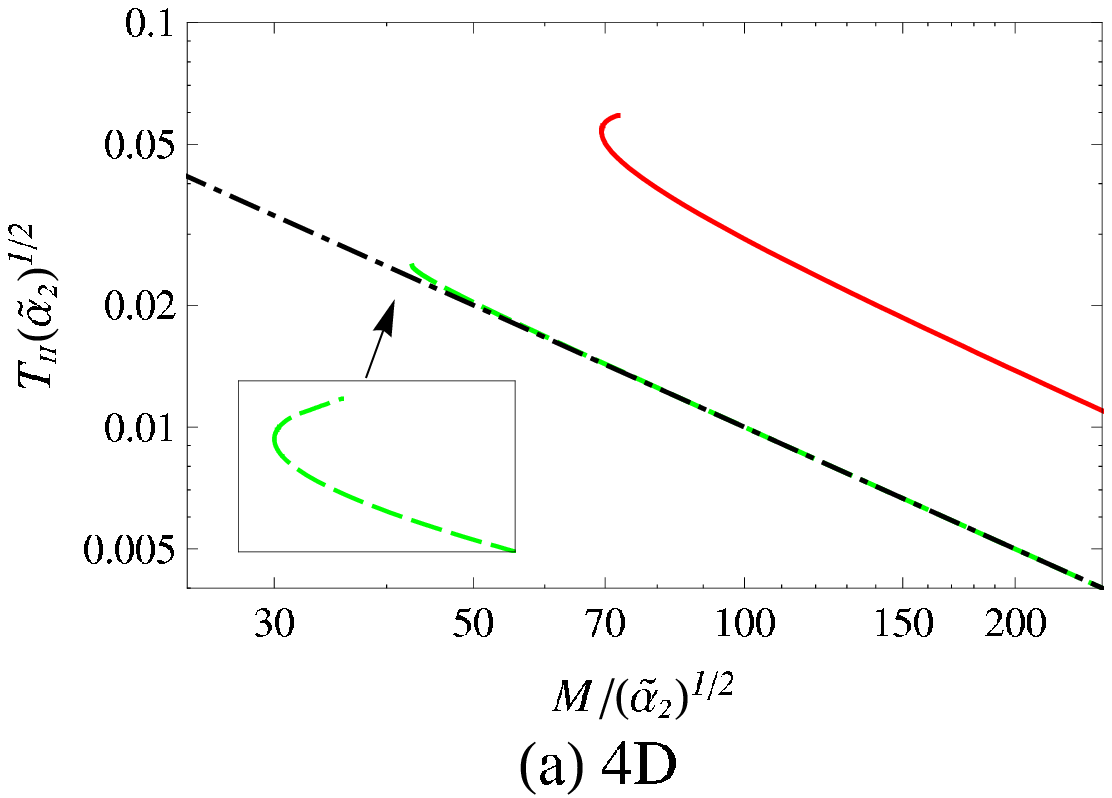}~~~
\includegraphics[width=58mm]{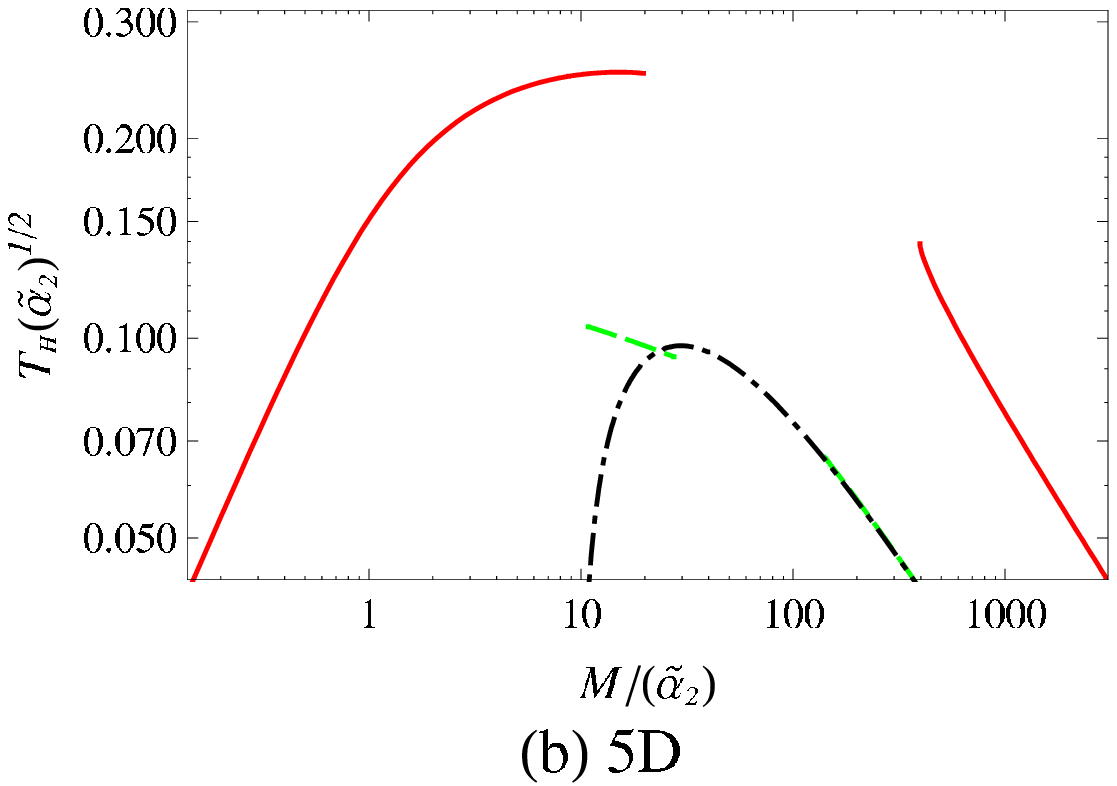}~~~
\includegraphics[width=58mm]{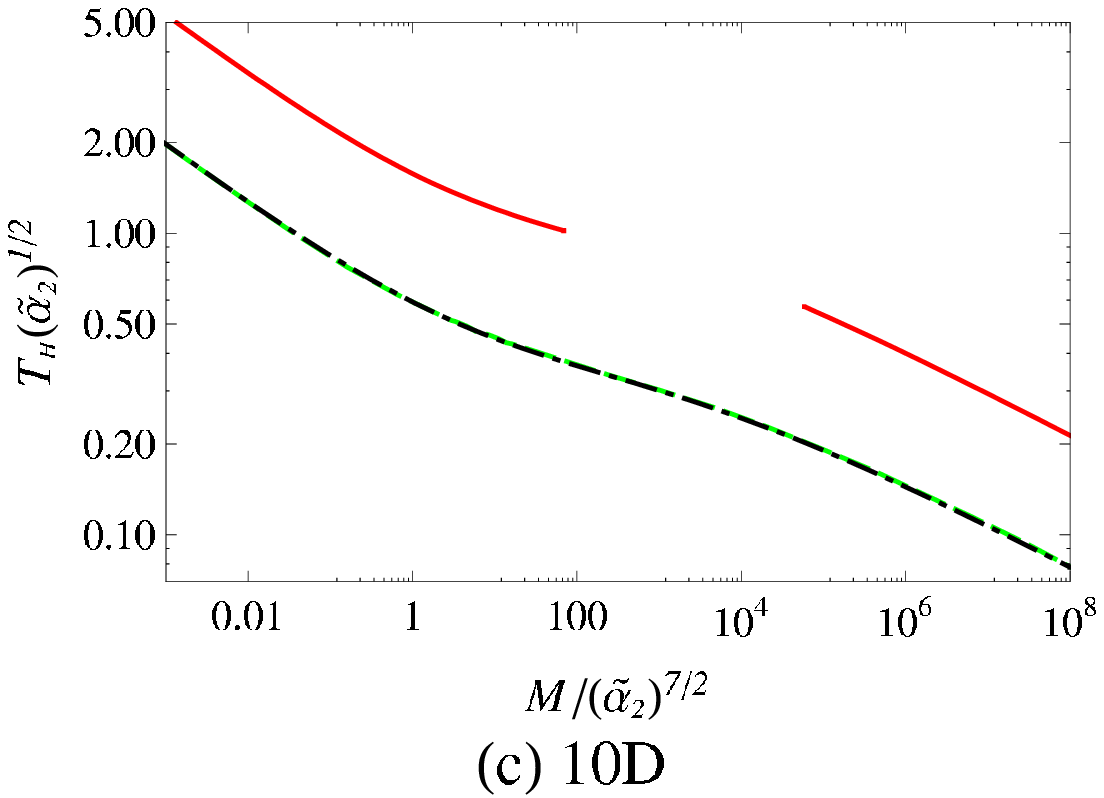}
\caption{The temperature of
black holes with respect to the gravitational mass.
The solid (red), dashed (green), and  dot-dashed (black) lines
are for the DEGB, TEGBD, and EGB models.
}
\label{temperature}
\end{figure}

\end{widetext}

In ten dimensions, the temperature decreases monotonically with respect to the mass
except near the turning point in the L-branch.
There is no maximum temperature just as in the TDEGB and EGB models.
In the L-branch of the DEGB model, however, we always find a turning point.

From this observation, we may conclude as follows:
In four dimensions, both the DEGB and TDEGB models predict almost the same.
When the black hole mass approaches the minimum value,
the temperature is still finite.
So the evaporation may not stop there.
We suspect that either it evolves into a naked singularity,
or it becomes time-dependent.

In five dimensions, however, two models give quite different predictions.
In the DEGB model, in the zero-mass limit, we find that the temperature
also vanishes. Then we expect that the black hole evaporate quietly.
On the other hand, in the TDEGB model,
no black hole exists below the minimum mass
$M_{\rm min}^{\rm (S)}$ and beyond the maximum temperature
$T_{\rm max}$.
When a back hole goes beyond this point via Hawking evaporation,
we will find a naked singularity or a time-dependent black hole
spacetime.

If the spacetime dimension is higher than five,
two models will give the similar fate, i.e.,
a black hole evaporates violently because the temperature diverges
in the mass-zero limit.

%%%%%%%%%%%%%%%%%%%%%%%%%%%%%%%%%%%%%%%%%%%%%%%%%%%%%%%%%%%%%%%%%%
%%%%%%%%%%%%%%%%%%%%%%%%%%%%%%%%%%%%%%%%%%%%%%%%%%%%%%%%%%%%%%%%%%
%%%%%%%%%%%%%%%%%%%%%%%%%%%%%%%%%%%%%%%%%%%%%%%%%%%%%%%%%%%%%%%%%%
\section{Concluding remarks}
\label{sec9}
%%%%%%%%%%%%%%%%%%%%%%%%%%%%%%%%%%%%%%%%%%%%%%%%%%%%%%%%%%%%%%%%%%
%%%%%%%%%%%%%%%%%%%%%%%%%%%%%%%%%%%%%%%%%%%%%%%%%%%%%%%%%%%%%%%%%%
%%%%%%%%%%%%%%%%%%%%%%%%%%%%%%%%%%%%%%%%%%%%%%%%%%%%%%%%%%%%%%%%%%
We summarize our results in Table \ref{table_4}.
The main difference in the DEGB model
from the TDEGB model
 is that the existence of a turning point in
five and higher dimensions
 and a zero-mass black hole in five dimensions.
The Hawking temperature in the five-dimensional DEGB model vanishes
at the zero-mass limit, but that in the TDEGB model
is finite.
The  DEGB model also gives a maximum
temperature in five dimensions. It may suggest that
the DEGB model is  better than the truncated one.
In fact, the maximum temperature is
given by $T_{\rm max}\approx 0.251938 \ta_2^{-1/2}
= 0.290913 \alpha'^{-1/2}$
at $M= 15.0506 \ta_2$, which is naively consistent with
the result given by the perturbative approach
($T_{\rm max}\sim 0.1 \alpha'^{-1/2}$)~\cite{Callan}.

We also include the result in the case of $\gamma=1/2$~\cite{ohta_torii1}.
The result in that model is almost qualitatively similar to
our TDEGB model except for the five dimensions.
In five dimensions, the result is the same as the Schwarzschild
black hole rather than that in our TDEGB or in the EGB model,
although we do not know the reason.

In this paper we consider only the asymptotically flat spacetime.
The asymptotically nonflat spacetimes, however, are also
important.
The asymptotically anti-de Sitter spacetime
is  especially interesting in the context of 
anti-de Sitter/conformal field theory (AdS/CFT) correspondence.
AdS/CFT correspondence is a widely-believed conjecture which suggests that
 there exists a duality between bulk gravity and boundary conformal theory.
Taking account into some quantum effects, i.e.,
the higher curvature correction terms,
one may examine a strong coupling region via AdS/CFT.
It may provide another confirmation for the conjecture.
The gravity duals of
Gauss-Bonnet gravity with nontrivial dilaton field
was studied in~\cite{Cai1}.
The asymptotically AdS spacetime in the TDEGB models
were also analysed~\cite{ohta_torii2,ohta_torii3}.
Since we find some important difference between
the DEGB and TDEGB models in this paper,
it is also interesting to analyse the asymptotically nonflat spacetimes
in the DEGB theory, which is under study.

\begin{widetext}

\begin{table}[ht]
\caption{The comparison between the DEGB and
TDEGB models with the EGB model and
Schwarzschild black hole as references.
In the four-dimensional TDEGB model,
we find a cusp instead of a turn-around smooth curve.}
\begin{center}
\begin{tabular}{|c||c||c|c||c||c|c|}
\hline
$D$&~
& DEGB
&TDEGB
&TDEGB($\gamma=1/2$)
&EGB
&Schwarzschild
\\
\hline
\hline
4&mass range
&$M>M_{\rm min}$
&$M>M_{\rm min}$
&$M>M_{\rm min}$
&
&$M\geq 0$
\\
\cline{2-5}\cline{7-7}
&turning point
&yes
&``yes (cusp)"
&no
 &Schwarzschild&no
\\
\cline{2-5}\cline{7-7}
&$T_{\rm max}$
&finite
&finite
&finite
  &black hole&  $\infty$
\\
\cline{2-5}\cline{7-7}
&$T_H$ at $M_{\rm min}$
&finite
&finite
&finite
& & $\infty$
\\
\hline
\hline
5&mass range
&$M_{\rm max}^{\rm (S)}> M \geq 0 \,,$
&$M_{\rm max}^{\rm (S)}> M>M_{\rm min}^{\rm (S)}\,,$
&$M>M_{\rm min}$
&$M > M_{\rm min}$
&$M\geq 0$
\\
&
&$M>M_{\rm min}^{\rm (L)}$
&$M>M_{\rm min}^{\rm (L)}$
&
&
&
\\
\cline{2-7}
&turning point
&yes
&no
&no
&no
&no
\\
\cline{2-7}
&$T_{\rm max}$
& finite
& finite
& $\infty$
& finite
&  $\infty$
\\
\cline{2-7}
&$T_H$ at $M_{\rm min}$
& zero
& finite
&  $\infty$
& zero
&  $\infty$
\\
\hline
\hline
6$\sim$10
&mass range
&$M_{\rm max}^{\rm (S)}>M \geq 0 \,,$
&$M\geq 0$
&
&$M\geq 0$
&$M\geq 0$
\\
&
&$M>M_{\rm min}^{\rm (L)}$
&
&the same as
&
&
\\
\cline{2-4}\cline{6-7}
&turning point
&yes
&no
& our TDEGB
&no
&no
\\
\cline{2-4}\cline{6-7}
&$T_{\rm max}$
& $\infty$
& $\infty$
&$(\gamma=\sqrt{2/(D-2)})$
& $\infty$
& $\infty$
\\
\cline{2-4}\cline{6-7}
&$T_H$ at $M_{\rm min}$
& $\infty$
& $\infty$
&
& $\infty$
& $\infty$
\\
\hline
\end{tabular}
\label{table_4}
\end{center}
\end{table}

%%%%%%%%%%%%%%%%%%%%%%%%%%%%%%%%%%%%%%
%%%%%%%%%%%%%%%%%%%%%%%%%%%%%%%%%%%%%%
\acknowledgments
%%%%%%%%%%%%%%%%%%%%%%%%%%%%%%%%%%%%%%
%%%%%%%%%%%%%%%%%%%%%%%%%%%%%%%%%%%%%%

N.O. would like to thank T. Torii for valuable discussions.
This work was partially supported
by the Grant-in-Aid for Scientific Research
Fund of the JSPS (Nos.19540308 and 20540283) and for the
Japan-U.K. Research Cooperative Program,
and by the Waseda University Grants for Special Research Projects.
%~~\\

%\newpage~~

\end{widetext}

\appendix
%\begin{widetext}
%%%%%%%%%%%%%%% %%%%%%%%%%%%%%%%%%%%%%%%%%%%%%%%%%
%%%%%%%%%%%%%%%%%%%%%%%%%%%%%%%%%%%%%%%%%%%%%%%%%%%
%%%%%%%%%%%%%%%%%%%%%%%%%%%%%%%%%%%%%%%%%%%%%%%%%%%
\section{Black hole in the Einstein-Gauss-Bonnet theory}
\label{EGB}
%%%%%%%%%%%%%%% %%%%%%%%%%%%%%%%%%%%%%%%%%%%%%%%%%
%%%%%%%%%%%%%%%%%%%%%%%%%%%%%%%%%%%%%%%%%%%%%%%%%%%
%%%%%%%%%%%%%%%%%%%%%%%%%%%%%%%%%%%%%%%%%%%%%%%%%%%

In this appendix, we summarize the properties of a black hole in the Einstein
Gauss-Bonnet theory. The action is
\begin{eqnarray}
{\cal S}_{\rm EGB} =\frac{1}{2\k_{D}^2} \int d^D \! x \sqrt{-g}
\biggl(R +  \a_{2}  R_{GB}^2 \biggr).
\label{EGBaction}
\end{eqnarray}
We find the field equations by setting the dilaton field $\phi=0$,
and can reduce them as
\begin{align}
&\left[ r^{D-3} \left(k-f(r) \right)
+\a_2 (D-3)_4 r^{D-5} \left(k-f(r) \right)^2 \right]'=0, \nn
& \quad \delta'(r) =0.
\end{align}
In four dimensions, the Gauss-Bonnet term does not give any contribution
to the solution.
We have just a Schwarzschild black hole, i.e., $f(r)=k-2\mu/r$.
For $D\geq 5$, we find two branches of the solutions as follows:
\begin{align}
f(r)=&f_{\pm}(r)
\nn
:=& k + \frac{ r^2}{2(D-3)_4\a_2} \left( 1
 \mp \sqrt{1 + {8  (D-3)_4\a_2 \mu \over r^{D-1}  }} \right), \nn
 \d(r)=&0
\,,
\end{align}
%\end{widetext}
where $\mu$ is an integration constant, which is related to the
gravitational mass $M$ as $(D-2)A_{D-2}\, \mu =  \kappa_D^2 M $.
The asymptotic behaviour or the weak coupling limit,
i.e., $\ta_2 \mu/r^{D-1} \ll 1$, gives
 \begin{align}
f_{+}(r) &\rightarrow  k - \left[{2\kappa_D^2 \over (D-2)A_{D-2}}\right]
 {M\over r^{D-3}},\nn
f_{-}(r) &\rightarrow  k + \left[{2\kappa_D^2 \over (D-2)A_{D-2}}\right]
 {M\over r^{D-3}}  + \frac{(D-2) r^2}{(D-4) \ta_2}.
\end{align}
The former is an asymptotically flat spacetime,
while the latter is an asymptotically anti-de Sitter
spacetime~\cite{footnote2}.
The black hole mass in the asymptotically flat case is given by
\begin{align}
\bar M &:=\kappa_D^2M
\nn
&= {(D-2)A_{D-2} \over 2} r_H^{D-3}\left[ 1 + {2(D-4) \ta_2 \over (D-2) r_H^2}
 \right]
\,,
\end{align}
and the Hawking temperature is
\begin{align}
T_{H}=
\frac{\left[ (D-2)_3 r_H^2 +  (D-4)_5 \ta_2 \right]}
{4 \pi r_H \left[ (D-2) r_H^2 + 2(D-4) \ta_2 \right]}
\,.
\end{align}
The entropy is given by the Wald's formula as
\begin{align}
S_{{\rm EGB}} = \frac{A_H}{4} \lh 1 +\frac{2 \ta_2}{r_H^2} \rh,
\end{align}

In four dimensions, it is just a Schwarzschild spacetime. There is nontrivial
contribution in the entropy from the Gauss-Bonnet term.
Then we find
\begin{eqnarray}
M_{\rm min}=0\,,~{\rm and}~~~ S_{\rm min}=2\pi\ta_2,
\end{eqnarray}
at $r_H=0$, when the temperature diverges ($ T_{\rm max}=\infty$).

In five dimensions, we find the black hole mass and the Hawking temperature as
\begin{align}
M = \pi^2 \left( 3 r_H^2 + 2 \ta_2 \right) \\
T_{H}= \frac{ 3 r_H } {2 \pi \left[ 3 r_H^2 + 2 \ta_2 \right]}
\,.
\end{align}
Then we find
\begin{eqnarray}
M_{\rm min}=2\pi^2 \ta_2\,,~ S_{\rm min}=0\,,~ T_{\rm min}=0,
\end{eqnarray}
at $r_H=0$.
We also find the maximum temperature as
\begin{eqnarray}
T_{\rm max}={\sqrt{6}\over 8\pi}\ta_2^{-1/2},
\end{eqnarray}
at $r_H=\sqrt{2 \tilde{\alpha_2}/3} $ ($M_{\rm max}=4\pi^2 \ta_2$).

For dimensions higher than five, we find
\begin{eqnarray}
M_{\rm min}=0\,,~{\rm and}~~~S_{\rm min}=0,
\end{eqnarray}
at $r_H=0$, when the temperature diverges ($T_{\rm max}=\infty$).

~~\\

% \newpage

%%%%%%%%%%%%%%%%%%%%%%%%%%%%%%%%%%%%%%%%%%%%%%%%%%%%%%%%%%%%%%%%%%%%
%%%%%%%%%%%%%%%%%%%%%%%%%%%%%%%%%%%%%%%%%%%%%%%%%%%%%%%%%%%%%%%%%%%%
%%%%%%%%%%%%%%%%%%%%%%%%%%%%%%%%%%%%%%%%%%%%%%%%%%%%%%%%%%%%%%%%%%%%

\begin{thebibliography}{99}
\bibitem{footnote1}
This scalar field $\phi$ does not give a proper
kinetic term.
Rescaling it as
$\varphi=\sqrt{2}\kappa_D\, \phi$,
$\varphi$ has a proper kinetic term.
\bibitem{footnote2}
{If $\ta_2<0$, we find an asymptotically de Sitter spacetime.}
\end{thebibliography}
\end{document}